\documentclass{agujournal}
\journalname{JGR-Space Physics}

\usepackage{siunitx}

\DeclareSIUnit{\AU}{AU}
\DeclareSIUnit{\sol}{sol}
\DeclareSIUnit{\solarradii}{R_{\mbox{$\odot$}}}

\begin{document}
	
	\title{Using Forbush decreases to derive the transit time of ICMEs propagating from 1 AU to Mars}
	\authors{Johan L. Freiherr von Forstner\affil{1},
		 Jingnan Guo\affil{1},
		 Robert F. Wimmer-Schweingruber\affil{1},
		 Donald M. Hassler\affil{2, 3},
		 Manuela Temmer\affil{4},
		 Mateja Dumbovi\'{c}\affil{4},
		 Lan K. Jian\affil{5, 6},
		 Jan K. Appel\affil{1},
		 Ja\v{s}a \v{C}alogovi\'{c}\affil{7},
		 Bent Ehresmann\affil{2},
		 Bernd Heber\affil{1},
		 Henning Lohf\affil{1},
		 Arik Posner\affil{8},
		 Christian T. Steigies\affil{1},
		 Bojan Vr\v{s}nak\affil{7},
		 and Cary J. Zeitlin\affil{9}
	}
	\affiliation{1}{Institute of Experimental and Applied Physics, University of Kiel, Germany}
	\affiliation{2}{Southwest Research Institute, Boulder, Colorado, USA}
	\affiliation{3}{Institut d'Astrophysique Spatiale, University Paris Sud, Orsay, France}
	\affiliation{4}{Institute of Physics, University of Graz, Austria}
	\affiliation{5}{University of Maryland, College Park, Maryland, USA}
	\affiliation{6}{NASA Goddard Space Flight Center, Greenbelt, MD, USA}
	\affiliation{7}{Hvar Observatory, Faculty of Geodesy, University of Zagreb, Croatia}
	\affiliation{8}{NASA Headquarters, Washington, DC, USA}
	\affiliation{9}{Leidos, Houston, Texas, USA}

\correspondingauthor{J. Guo}{guo@physik.uni-kiel.de}

\begin{keypoints}
    \item The interplanetary propagation of 15 CMEs is studied based on a cross-correlation analysis 
    of  Forbush decreases at 1 AU and Mars.
    \item The speed evolutions of the ICMEs are derived from observations, indicating that most of them are slightly 
    decelerated even beyond 1 AU.
    \item Model-predicted ICME arrival times at Mars could be improved by using ICME parameters measured at 1 AU.
\end{keypoints}

\begin{abstract}
    The propagation of 15 interplanetary coronal mass ejections (ICMEs) from Earth's orbit (\SI{1}{\AU}) to Mars ($\sim 
\SI{1.5}{\AU}$) has been studied with their propagation speed estimated from both measurements and 
simulations. The enhancement of magnetic fields related to ICMEs and their shock fronts cause the so-called Forbush 
decrease, which can be detected as a reduction of galactic cosmic rays measured on-ground. We have used galactic cosmic 
ray (GCR) data from in-situ measurements at Earth, from both STEREO A and B as well as GCR measurements by the 
Radiation Assessment Detector (RAD) instrument onboard Mars Science Laboratory (MSL) on the surface of Mars. A set of 
ICME events has been selected
during the periods when Earth (or STEREO A or B) and Mars locations were nearly aligned on the same
side of the Sun in the ecliptic plane (so-called opposition phase). Such lineups allow us to estimate the ICMEs'
transit times between \SIrange[range-phrase={~and~}]{1}{1.5}{\AU} by estimating the delay time of 
the corresponding Forbush decreases measured at each location. 
We investigate the evolution of their propagation speeds before and after passing Earth's orbit and find that the 
deceleration of ICMEs due to their interaction with the ambient solar wind may continue beyond \SI{1}{\AU}. We also 
find a substantial variance of the speed evolution among different events revealing the dynamic and diverse nature of 
eruptive solar events. Furthermore, the results are compared to simulation data obtained from two CME propagation 
models, namely the Drag-Based Model and ENLIL plus cone model.
\end{abstract}

\section{Introduction}

It is currently well accepted that Coronal Mass Ejections (CMEs), magnetized plasma clouds expelled from the Sun, may 
have severe impact on Earth, robotic missions on other planets, as well as spacecraft electronics. A better 
understanding of the interplanetary propagation of CMEs is very important to gain a deeper understanding of the 
heliosphere and the Sun itself, and to improve space weather forecasting.

ICMEs are regularly observed using both remote sensing images (coronagraph and heliospheric imaging 
instruments) and \textit{in situ} measurements of plasma and magnetic field quantities 
\replaced{[e.g. Wimmer-Schweingruber et al., 2006]}{\citep[e.g.][]{Richardson1995, Wimmer-Schweingruber-2006}}. Another 
common 
\textit{in situ} 
method to detect ICMEs employs the observation of Forbush decreases \replaced{[Forbush, 1937; Cane, 2000, and 
references 
therein]}{(first observed by \citet{Forbush-1937,Hess-1937} and also studied by e.g. 
\citet{Lockwood1971,Burlaga1985,Cane-2000,Kumar2014,Zhao2016})} in measurements 
of galactic cosmic rays (GCR) caused by the magnetic field structure embedded in the ICME passing by.

ICMEs can not only consist of the magnetized ejecta (which, depending on its geometry, can also be called 
``magnetic cloud'' or ``flux rope''), but in many cases also drive an interplanetary shock in front of them, separated 
by a turbulent sheath region. Forbush decreases can occur during the passage of the sheath region (after the shock 
arrival) as well the ejecta, which is described to be the cause of a two-step structure e.g. by \citet{Cane-2000}. 
However, recent studies such as \citet{Jordan-2011} and \citet{MasiasMeza-2016} have found that even though the ejecta 
is effective 
at decreasing the GCR intensity, an ICME with a shock does not necessarily produce a clear two-step structure in the FD 
and that the shock arrival is much more likely to produce an abrupt drop in the GCR intensity than the ejecta. 
\added{When multiple CMEs are ejected from the Sun in a short period of time, they can interact with each other during 
their propagation and form complex structures, which also affects the corresponding Forbush decrease 
\citep[e.g.][]{Maricic2014}.}

The interplanetary propagation of ICMEs is strongly influenced by the ambient solar wind. This leads to either a 
deceleration or acceleration 
depending on the relative speed of the ICME to the ambient solar wind speed 
\citep[e.g.][]{Gopalswamy-2001,Vrsnak-2004,Vrsnak-2007}. As most CMEs launched from the 
Sun are faster than the ambient solar wind, this more often results in deceleration rather than 
acceleration. With a large amount of imaging and \textit{in situ} instruments available 
on spacecraft especially near Earth's orbit, extensive studies of the evolution of CMEs during their 
eruption at the Sun and their propagation up to \SI{1}{\AU} have been carried out. Heliospheric imaging instruments, 
for example on board the two \textit{Solar Terrestrial Relations 
Observatory} (STEREO) spacecraft, allow a continuous tracking of ICMEs up to \SI{1}{\AU} 
\citep[e.g.][]{Lugaz-2012,Moestl-2014,Wood-2017}.
Additionally, spacecraft such as \textsc{Ulysses} 
\citep[e.g.][]{Wang-2005,Jian-2008} and \textsc{Voyager} \citep[e.g.][]{Liu-2014} have provided ICME observations at 
locations in the outer solar system. Based on the results from \citet{Wang-2005} and their own studies of ICMEs seen 
at Mercury and Earth, \citet{Winslow-2015} stated that the deceleration of most ICMEs should cease at approximately 
\SI{1}{\AU}.

With the \textsc{Curiosity} rover of NASA's \textit{Mars Science Laboratory} (MSL) mission \citep{Grotzinger-2012}, 
another device capable of 
registering Forbush decreases is available on the surface of Mars (at approximately \SI{1.5}{\AU}) since its landing on 
Aug 6, 2012. Its \textit{Radiation Assessment Detector} (RAD) instrument \citep{Hassler-2012} has been 
continuously measuring GCR particles on the surface of Mars since then. MSL/RAD was already used 
for observations of ICMEs through Forbush decreases, for example by \citet{Witasse-2017}.

In situ observations of ICMEs at Mars are also possible using instruments on the \textit{Mars Atmosphere 
and Volatile Evolution} (MAVEN) spacecraft which is in orbit around Mars. But it only arrived at Mars in 
September 2014, so a time period of two years after MSL's arrival can not be studied using MAVEN data. 
For this reason, we have not yet incorporated MAVEN data in this study, but we plan to do so in the 
future as the number of ICMEs observed by MAVEN increases. \added{A first study involving a comparison of ICME 
measurements at MAVEN and MSL/RAD can be found in \citet{Guo-inpreparation}}.

\begin{figure}[htb]
    \centering
    \includegraphics{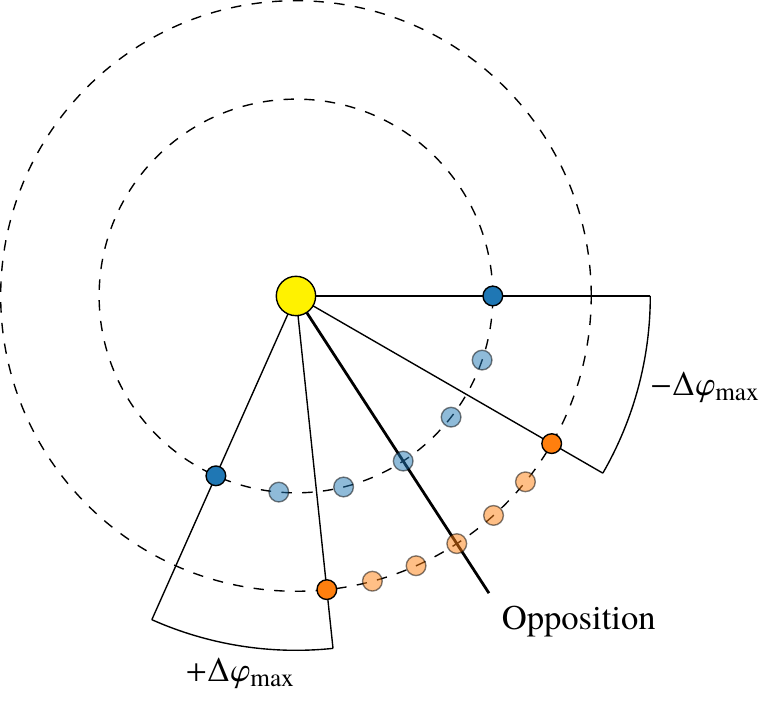}
    \caption{The opposition phases for this study are defined by the longitudinal separation of Earth (or STEREO) and 
    Mars being between $-\Delta \varphi_\text{max}$ and $+\Delta \varphi_\text{max}$. The bold diagonal line marks the 
    opposition itself.}
    \label{fig:earth-mars-conjunction}
\end{figure}

At times where Mars and either Earth or the STEREO A or B spacecraft have a low separation in their heliospheric 
longitudes, i.e. they nearly form a straight line with the Sun, we have a better chance of observing the same ICMEs at 
both \SI{1}{\AU} and Mars 
using \textit{in situ} data. These times are the oppositions of Mars observed from Earth and the STEREO spacecraft, 
respectively. We define an \textit{opposition phase} to be the period where the absolute value of the longitudinal 
separation $\Delta \varphi$ between Mars and Earth (or STEREO) is smaller than a fixed value $\Delta 
\varphi_\text{max}$, which for this study is set to \SI{30}{\degree}, keeping the probability that ICMEs are 
observed at both locations reasonably high, but at the same time not restricting the number of ICME candidates too 
much. The latitudinal separations between Earth, the STEREO spacecraft 
and Mars are generally only a few degrees at most and therefore not taken into account.
\citet{Yashiro-2004} found that the average angular width of CMEs is between 
\SIrange[range-phrase={~and~}]{47}{61}{\degree}, which supports that choosing $\Delta \varphi_\text{max} = 
\SI{30}{\degree}$ is reasonable.
Figure \ref{fig:earth-mars-conjunction} illustrates the opposition phases and the definition of $\Delta \varphi_\text{max}$.

These multi-spacecraft observations of ICMEs during the opposition phases allow us to determine ICMEs' travel times 
between the radial distances of \SI{1}{\AU} and $\sim \SI{1.5}{\AU}$ from the Sun. They can be used to compare the 
resulting transit speed with measurements at \SI{1}{\AU} to determine the amount of deceleration or acceleration.

To be able to derive the propagation time of an ICME between the two observation locations, we assume that the 
same part of the ICME is observed at both places, or alternatively that the ICME's shape has a sufficient amount of 
radial symmetry between the two longitudes where it is observed. The probability that this assumption holds true is 
obviously higher for smaller longitudinal separations of the two observers, which is another reason why we chose a 
small angle for $\Delta \varphi_\text{max}$. A sophisticated study of the ICMEs' shapes could only in theory be done with a 
significantly higher number of observation locations or using 3-D reconstruction techniques based 
on stereoscopic imaging techniques, 
where the former drastically reduces the amount of ICME candidates and the latter can currently only be done up to 
approximately \SI{1}{\AU}, e.g. with the heliospheric imagers at both STEREO spacecraft 
\citep[e.g.][]{Liu-2010}.

\section{Methods and Data}

\subsection{Data}

Table \ref{tab:opposition-periods} shows the opposition periods between \textsc{Curiosity}'s landing in August 2012 and 
the end of 2016, as defined in Figure \ref{fig:earth-mars-conjunction}. Oppositions of Earth and Mars are included as 
well as those with the STEREO spacecraft.

\begin{table}[ht]
	\centering
	\begin{tabular}{rccc}
		\hline
		& \multicolumn{3}{c}{Date}\\
		\cline{2-4}
		Opposition type & Start & Opposition & End \\
		\hline
		STEREO B \& Mars & 2012-08-22 & 2012-11-28 & 2013-02-05 \\
		STEREO A \& Mars & 2013-05-21 & 2013-07-19 & 2013-09-12 \\
		Earth \& Mars    & 2014-02-13 & 2014-04-08 & 2014-06-10 \\
		Earth \& Mars    & 2016-03-20 & 2016-05-22 & 2016-08-05 \\
		\hline
	\end{tabular}
	\caption{Opposition periods considered for this study. The start and end dates of the $\pm \SI{30}{\degree}$ 
		periods and the actual date of the opposition are given.}
	\label{tab:opposition-periods}
\end{table}

We used the ICME list by \citet{Richardson-Cane-2010,data:richardson-cane-icme-list} as a basis for finding the 
ICME-caused Forbush 
decreases at Earth, and a similar list by \citet{Jian-2013} for ICMEs at STEREO A and B.

Communication with the STEREO B spacecraft was lost on October 1, 2014 and a recovery attempt in summer 2016 was 
not successful. Therefore, data from its 2015 opposition with Mars is not available. Additionally, because of the solar 
conjunction in 2015, PLASTIC onboard STEREO A was turned off and there is no plasma data for the second STEREO A \& 
Mars opposition phase. For these reasons, we excluded the two 2015 opposition phases, leaving us with the four 
opposition periods to investigate in this study.

For the two oppositions of Earth and Mars, we retrieved count rate data from the Neutron Monitor Database 
(\texttt{http://nmdb.eu}). We chose the South Pole neutron monitor (abbreviated as SOPO), which has a low cutoff 
rigidity (with an effective vertical cutoff rigidity of \SI{0.1}{\giga\volt}) due to its geographic location. This was 
then used together with the RAD dose rate data to apply the 
cross-correlation method, which will be described in section \ref{subsec:cross-correlation}.

For the STEREO oppositions, we replaced the neutron monitor count rates with measurements from the High Energy 
Telescope (HET) instruments available on both STEREO spacecraft \citep{vonRosenvinge-2008}, which measure the flux of 
high-energy charged particles.
While \textit{in situ} observations of ICMEs at the STEREO spacecraft are also possible using 
magnetometer and plasma data (as has been used to identify ICMEs in the lists employed in our study), Forbush decreases 
in the HET data allow for a more direct comparison to the RAD data at Mars.

The publicly available HET data includes measurements of protons with kinetic energies between 
\SIrange[range-phrase={~and~}]{13.6}{100}{\mega\electronvolt} and electrons between 
\SIrange[range-phrase={~and~}]{0.7}{4.0}{\mega\electronvolt}, with each of these ranges subdivided into 
multiple 
energy bins.
We chose the \SIrange[range-phrase={~to~}]{23.8}{100}{\mega\electronvolt} proton range, which appeared to show the 
Forbush decreases reasonably well for this study. In some cases (event numbers: 1, 2, 4, 5, 7 and 9), we chose to only 
use the highest energy 
channel (\SIrange[range-phrase={~to~}]{60}{100}{\mega\electronvolt}) instead because there were solar energetic 
particles (SEP) coinciding with the ICME arrival at STEREO, resulting in a much higher particle flux instead 
of FDs in the lower HET channels.

\subsection{RAD data and compensating for the diurnal variations}
\label{subsec:diurnal}
RAD/MSL is an energetic particle detector and it has been carrying out radiation measurements on the surface of Mars 
since the landing of MSL in August 2012 \citep{hassler2014, ehresmann2014, Rafkin-2014, koehler2014, guo2015modeling, 
wimmer2015, Guo-2017}. On the surface of Mars, RAD measures a mix of primary GCRs or SEPs and secondary particles 
generated in the atmosphere including both charged and neutral particles. Due to the shielding of the atmosphere, such 
particles are mostly equivalent to primary GCR/SEP with energies larger than $\sim$ \SI{100}{\mega\electronvolt\per 
nuc}. The radiation dose 
rates contributed by surface particles are measured in two detectors --- a silicon detector and a plastic scintillator 
--- and the latter has better statistics due to a larger geometric factor and is a very good proxy for studying GCR 
fluence and its temporal variations.

RAD's GCR dose rate measurements on the surface of Mars show a considerable amount of periodic variation (about 
$\pm\SI{5}{\percent}$) with a frequency of \SI{1}{\sol} and its harmonics, which is caused by the variation of 
temperature and therefore 
atmospheric pressure during the course of the Martian day. This effect was analyzed by \citet{Rafkin-2014} and its 
intensity varies for different fluxes of primary and secondary GCR particles.

\citet{Guo-2017} found that the magnitude of this diurnal effect is not constant, but rather influenced by the solar modulation of the primary GCRs, direct subtracting of the pressure effect during an FD event is therefore not feasible.
To reliably detect Forbush decreases in this data, we process the data using a notch filter \citep{parks1987digital} that significantly reduces 
the diurnal variations in the data, but keeps other influences --- such as Forbush decreases --- intact.
A more detailed description of the implementation of this method is shown in \citet{Guo-inpreparation}.

\subsection{Cross-correlation analysis}
\label{subsec:cross-correlation}
We assume that the travel time of the ICMEs between \SI{1}{\AU} and Mars corresponds to the delay time between the 
onset of Forbush decreases detected at these two locations. To determine this delay, we use a method based on the 
cross-correlation function (CCF), assuming that Forbush decreases at \SI{1}{\AU} and Mars from the same ICME should 
have similar characteristics, such as being a one- or two-step decrease. An advantage of this method is that it 
allows to determine the travel time without 
needing to define exact onset times at both Earth and Mars, which can be difficult when the Forbush decrease is weak 
and/or rather complex.

For the analysis, a $\pm\SI{1}{\sol}$ window (a sol is a solar day on Mars, $\SI{1}{\sol} \approx 
\SI{24}{\hour}\ \SI{40}{\minute}$) around the given ICME onset time at 
\SI{1}{\AU} $t_{\SI{1}{\AU}}$ is selected from the GCR data at \SI{1}{\AU}, which includes a Forbush decrease at this 
time. The rather small window makes sure that we only compare the actual decrease, so that a difference in the 
following recovery period should not affect the results.
The normalized cross-correlation function of the \SI{1}{\AU} data with the filtered RAD dose rate 
data (see details in section \ref{subsec:diurnal}) is then calculated in this window. It is a measure for the 
correlation between the two datasets when one is 
shifted in time by a lag $\tau$. For discrete measurements $f[m]$ and $g[m]$, the normalized cross-correlation function 
is defined as

\begin{equation}
(f \star g)[n] \doteq \frac{1}{m_\text{max} - m_\text{min}}\sum_{m = m_\text{min}}^{m_\text{max}} f'[m]\ g'[m+n],
\end{equation}
where the lag $\tau$ is represented by a number of data points $n$, the range $[m_\text{min}, m_\text{max}]$ is the 
aforementioned $\pm\SI{1}{\sol}$ window, and the normalized functions $f', g'$ are defined as

\begin{equation}
\begin{aligned}
f'[m] \doteq \frac{f[m] - \overline{f}}{\sigma_f},\\
g'[m] \doteq \frac{g[m] - \overline{g}}{\sigma_g}.
\end{aligned}
\end{equation}
where $\sigma_f$ and $\sigma_g$ are standard deviations of $f[m]$ and $g[m]$ in the range $[m_\text{min}, m_\text{max}]$, 
respectively.

The value of $\tau$ where $(f \star g)$ assumes its maximum in a reasonable range  $\tau \in [0, \Delta 
t_\text{max}]$ is considered to be the ICME's travel time $T$ between \SI{1}{\AU} and Mars. We fit the 
cross-correlation function's peak with a Gaussian distribution to estimate the error of $T$.

\begin{figure}[htb]
	\centering
	\includegraphics{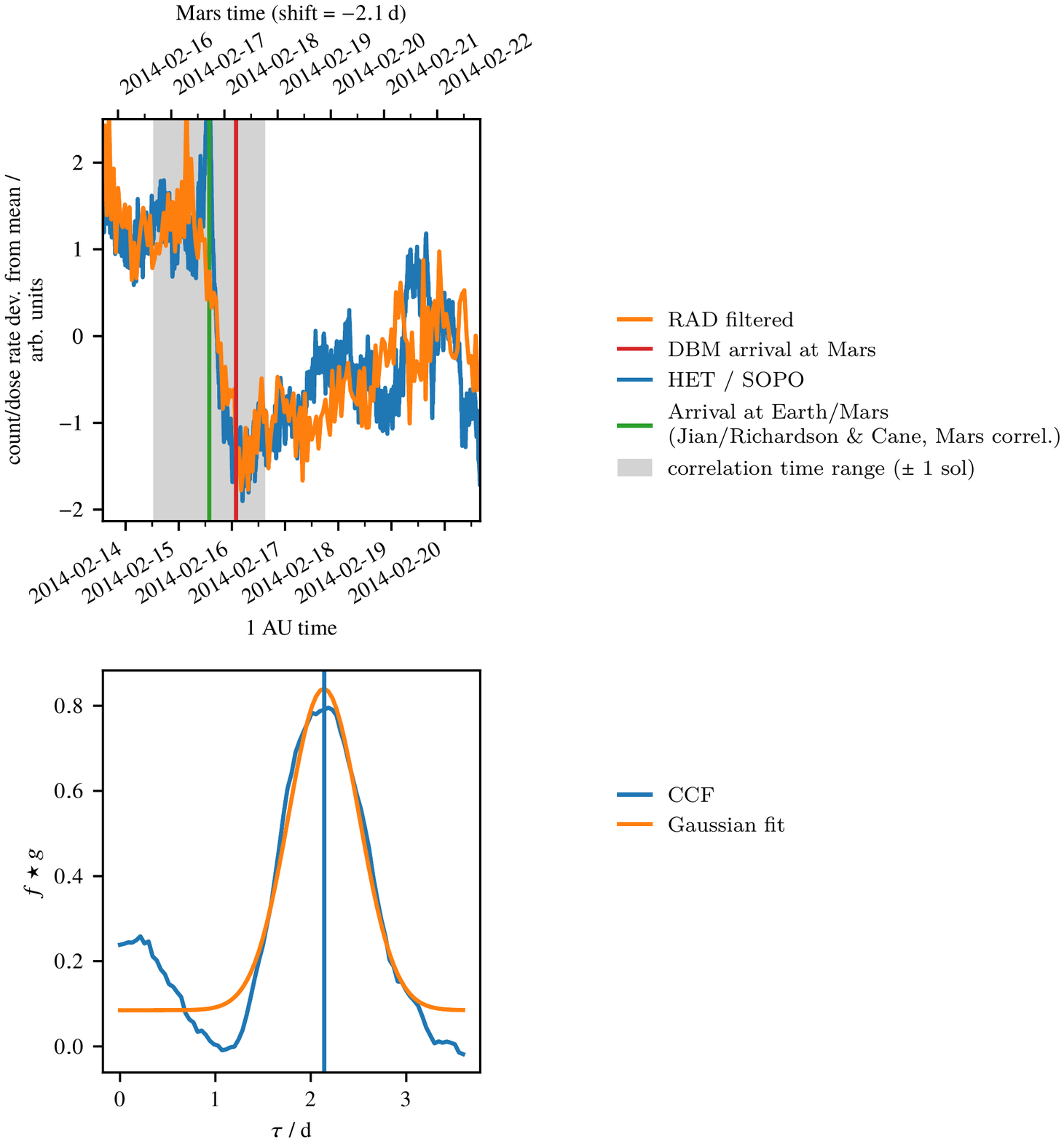}
	\caption{Example of an application of the cross-correlation method. The top panel shows count rate data from the 
		South Pole neutron monitor with the ICME disturbance time from the 
		\citet{data:richardson-cane-icme-list} ICME list marked by the green bar. In the same panel, the RAD dose 
		rate data (filtered using the method described in section \ref{subsec:diurnal} and shifted back 
		in time by the best-fitted CCF lag time) is shown, together with the onset time 
		calculated using the DBM model (red bar). The bottom panel shows the cross-correlation function (CCF) of the 
		two datasets 
		plotted over the time lag $\tau$, fitted with a Gaussian function to obtain the estimated 
		travel time $T$ and its uncertainty. The window used for calculating the CCF is displayed in the upper panel 
		with a light gray color.
		In this example, the resulting ICME travel time is \SI{2.14+-0.37}{\day}, which is slightly shorter than the 
		\SI{2.6}{\day} 
		calculated using the DBM model (explained later in section \ref{subsec:models}).}
	\label{fig:correlation_example}
\end{figure}

Figure \ref{fig:correlation_example} shows an example of an application of the cross-correlation method applied to the 
ICME that arrived at Earth on 2014-02-15. The implication of these results will be discussed in section 
\ref{sec:results}.

The GCR data in Figure \ref{fig:correlation_example} is scaled so that the correlation between the two datasets is more clearly 
shown. 
Specifically, we subtract the mean value in the shown timerange from the measurements and then divide the results by 
their standard deviations. For some events at the STEREO spacecraft, we adjusted the scaling of the HET flux rate data 
manually as the calculation of the mean and standard deviation was affected by strong increases in the data related to 
SEP events shortly before or after the ICME arrival. This was done by calculating the mean and standard deviation in a 
smaller $\pm\SI{16}{\hour}$ period around $t_{\SI{1}{\AU}}$ instead of the whole range of the plot. Additionally, 
in one case (event 1) we needed to decrease the size of the window in which the correlation is calculated to 
$\pm\SI{0.75}{\sol}$ instead of $\pm\SI{1}{\sol}$ to make sure that the SEP event does not influence the result of the 
cross-correlation analysis.

Note that we are not comparing the magnitude of the Forbush decreases at \SI{1}{\AU} and 
Mars, which would be an interesting study in the future. However, it needs to be considered that both 
the neutron 
monitor measurements and RAD dose rate are influenced by the atmosphere and/or magnetosphere of two different planets, 
which makes the comparison more complicated than simply assessing the relative drop ratios in the two datasets.

\section{Results and Discussion}

\subsection{Results}
\label{sec:results}

In total, 43 ICMEs were observed during the four opposition periods, according to the Richardson/Cane \citep{Richardson-Cane-2010,data:richardson-cane-icme-list} and Jian \citep{Jian-2013} lists. 
However, not 
all of them caused visible Forbush decreases in our datasets at \SI{1}{\AU} and/or Mars --- probably 
because \textit{a)} FDs can be very weak in comparison to the background oscillations, e.g. due to low 
ICME speeds and/or magnetic field strengths, \textit{b)} the ICME missed one of the observation 
points, e.g. due to \textit{1)} the angular width of the ICME is not covering the longitudinal separation of 1 
AU and Mars observers (up to $\pm \SI{30}{\degree}$) and/or \textit{2)} a significant deviation of the propagation 
direction, \textit{c)} a gap occurring in the data at one of the observation locations or 
\textit{d)} a strong solar energetic particle (SEP) event seen at STEREO does not allow us to see the FD even when 
selecting only the highest energy channels.

Additionally, a considerable amount of ICMEs were ejected from the sun in quick succession and possibly
interacted or merged with each other during their propagation, which makes the cross-correlation analysis very 
difficult. One example for this is a series of 5 events in early February 2014 at Earth, where only the first 
event could be analyzed sufficiently well using the cross-correlation method as its distance to the others was larger 
and it had the strongest FD.

We therefore only kept the events in the study where the onset time from the list corresponded to a clear FD 
at \SI{1}{\AU} and where a convincing correspondence to a FD at Mars could be found using the cross-correlation 
analysis. For the 15 remaining events, we are most confident that the cross-correlation method picked up the Forbush 
decreases corresponding to the same ICME in the datasets at both locations.

In total, 14 events had no or only weak Forbush decreases at at least one of the observation locations causing a 
high uncertainty in the cross-correlation method results, 10 events were dropped due to a merging of multiple ICMEs, 3 
FDs at STEREO could not be seen due to a coinciding SEP event, and one event could not be analyzed due to a gap in the 
RAD data. A comparison of the speeds $v_\text{max}$ listed in the Richardson and Cane/Jian lists of the full set of 43 events 
to our selection of 15 events shows that both nearly have the same average value of \SI{476}{\kilo\meter\per\second} 
and \SI{475}{\kilo\meter\per\second}, respectively, so it seems that we did not select a set of particularly fast 
ICMEs.

The Richardson/Cane and Jian lists include arrival times for multiple ICME features: The disturbance arrival time 
(which refers to the arrival of a shock), the ICME plasma arrival time and the ICME end time. For the ICMEs 
where the disturbance arrival time was listed, we used it as the basis for the cross-correlation method because 
as explained in the introduction, the shock is most likely to cause the FD. For the remaining events, we used the ICME 
plasma arrival time. 

In fact, the choice of the onset time used at Earth hardly affects our study of the propagation time as it is only 
used to determine the position of the $\pm\SI{1}{\sol}$ window (which is sufficiently large in comparison to the onset 
time precision) in which the cross-correlation function is calculated. Nevertheless, in a few cases (events 11, 12 and 
13) the onset times 
were manually corrected ``by eye'' by amounts of up to a few hours to better reflect the beginning of the 
Forbush decrease in the \textit{in situ} data at Earth, which is not necessarily equal to the start of the disturbance 
or ICME start given in the lists.

Table \ref{tab:icmes1} shows the basic data and the results of the cross-correlation method for all the ICMEs in this 
study. Figures \ref{fig:icmes1} to \ref{fig:icmes5} in the appendix include the corresponding plots of the in situ data 
and CCF.

As explained in section \ref{subsec:cross-correlation}, due to the uncertainties in the data, the CCF was fitted with a Gaussian distribution to both enhance 
the detection of the maximum and obtain an estimation of the error. This method works 
reasonably well most of the time, but in some cases (e.g. events 9 and 14), the CCF shows a relatively wide peak, 
overlaid by one or multiple narrow peaks. Especially in these cases, the error might have been 
overestimated by the 
fit, which generally follows the wide peak.

For each ICME, the ratio $\overline{v}/v_{\SI{1}{\AU}}$ was calculated and shown Table \ref{tab:icmes1}, where $v_{\SI{1}{\AU}}$ 
is the measured maximum
speed of the ICME at \SI{1}{\AU}, obtained from the Richardson/Cane and Jian lists ($v_\text{max}$ column --- maximum solar 
wind speed during the passing of the ICME and shock/sheath), which presumably corresponds to the \replaced{speed of the 
leading edge}{propagation speed} of the shock (if present) or the ejecta; and
$\overline{v}$ is the average speed of the ICME between \SI{1}{\AU} and 
Mars calculated from the travel time obtained from the cross-correlation method and the radial distance $\Delta r$ 
between Earth (or the STEREO spacecraft) and Mars:
\begin{equation}
\overline{v} = \frac{\Delta r}{T_\text{correl}}
\label{eq:mean_speed}
\end{equation}

Additionally, if we assume that the acceleration $a$ of the ICME between \SI{1}{\AU} and Mars is constant, we can 
calculate it from the travel time $T_\text{correl}$ and the measured speed at \SI{1}{\AU} using the following 
considerations: With $v(t) = v_{\SI{1}{\AU}} + at$, the mean speed $\overline{v}$ (as given in equation 
\ref{eq:mean_speed}) can also be expressed as:
\[
\overline{v} = \frac{\int_{0}^{T_\text{correl}} v(t) \mathop{}\!\mathrm{d} t}{T_\text{correl}} = v_{\SI{1}{\AU}} + 
\frac{1}{2} a T_\text{correl}
\]
Equating this expression with the one from equation \ref{eq:mean_speed} and solving for $a$ gives:
\begin{equation}
a_{\SI{1}{\AU}, \text{Mars}} = 2\left( \frac{\Delta r}{T_\text{correl}^2} - \frac{v_{\SI{1}{\AU}}}{T_\text{correl}} 
\right)
\label{eq:acceleration}
\end{equation}
This value was also calculated for all events and included in Table \ref{tab:icmes1}. Similarly, the mean acceleration 
between a radial distance of \SI{21.5}{\solarradii} from the Sun and the arrival at \SI{1}{\AU}, $a_{\text{Sun}, \SI{1}{\AU}}$, was calculated using 
the speed at \SI{21.5}{\solarradii} obtained from the DONKI database (Section \ref{subsec:models}) and the travel 
time between those locations. In the case of ICME 2, the launch speed of \SI{480}{\kilo\meter\per\second} from 
the DONKI database was changed to the more reasonable value of \SI{960}{\kilo\meter\per\second} reported in the 
SOHO/LASCO CME catalog for the calculation of the acceleration. The DONKI value of \SI{480}{\kilo\meter\per\second} led 
to a negative, unphysical result for the drag parameter $\Gamma$ calculated in section \ref{subsec:dbm}. For the other 
events, the difference between the DONKI and SOHO/LASCO catalogs was much less significant. 

\begin{sidewaystable*}
    \centering
    \caption{Table of all the ICMEs examined in this study. The second column shows the spacecraft or planet at 
    \SI{1}{\AU} where the ICME was observed (STEREO A, STEREO B or Earth) and the third column contains the ICME 
    arrival time $t_{\SI{1}{\AU}}$ at this location according to the Richardson/Cane or Jian list (disturbance start 
    time if available, otherwise ICME start time). Column 4 states the 
    speed $v_{\SI{1}{\AU}}$, also taken from the Richardson/Cane and Jian lists ($v_\text{max}$). $\Delta r$ is the radial distance between Earth and Mars or the STEREO spacecraft and Mars, 
    respectively, at the time $t_{\SI{1}{\AU}}$. The next two columns include the CME launch speed 
    $v_{\text{launch}}$ used for simulation purposes and the average ambient solar wind speed $v_{\text{sw}}$ in the 3 
    days before the ICME arrival at \SI{1}{\AU}. $T_\text{correl}$ is the 
    estimated travel time obtained 
    from the cross-correlation method. The arrival time at Mars $t_\text{Mars}$ was calculated under the assumption 
    that $t_{\SI{1}{\AU}}$ is correct, and $\overline{v}$ is the average ICME speed between Earth orbit and 
    Mars calculated from $T_\text{correl}$ and $\Delta r$.
	The final column shows the ratio $\overline{v}/v_{\SI{1}{\AU}}$.
	The last row shows the average values (if applicable) for all events together.}
    \label{tab:icmes1}
    \begin{tabular}{@{\extracolsep{4pt}}cccS[table-format=3.0,round-mode=figures,round-precision=3]S[table-format=1.3,round-mode=figures,round-precision=3]S[table-format=3.0]S[table-format=3.0(3)]S[table-format=1.2(2)]cS[table-format=4.0(3)]S[table-format=1.2(2)]@{}}
\hline
& \multicolumn{6}{c}{Observations} & \multicolumn{4}{c}{Cross-correlation method results}\\ \cline{2-7}\cline{8-11}
{ICME} & {Obs S/C} & {$t_{\SI{1}{AU}}$} & {$v_{\SI{1}{\AU}}$} & {$\Delta r$} & {$v_{\text{launch}}$} & {$v_{\text{sw}}$} & {$T_{\text{correl}}$} & {$t_\text{Mars} = t_{\SI{1}{AU}} + T_{\text{correl}}$} & {$\overline{v}$} & {$\frac{\overline{v}}{v_{\SI{1}{\AU}}}$}\\
{} & & {(UTC)} & {/ \si{\kilo\meter\per\second}} & {/ \si{\AU}} & {/ \si{\kilo\meter\per\second}} & {/ \si{\kilo\meter\per\second}} & {/ \si{\day}} & {(UTC)} & {/ \si{\kilo\meter\per\second}} & {}\\
\hline
1 & STB & 2012-09-25 16:26 & 740.0 & 0.407 & 1056 & 389+-12 & 1.24+-0.23 & 2012-09-26 22:13 & 567+-103 & 0.77+-0.14\\
2 & STB & 2012-10-17 06:57 & 365.0 & 0.369 & 960 & 295+-5 & 2.05+-0.33 & 2012-10-19 08:16 & 311+-51 & 0.85+-0.14\\
3 & STB & 2012-10-25 19:10 & 435.0 & 0.357 & 380 & 297+-5 & 2.31+-0.66 & 2012-10-28 02:39 & 267+-76 & 0.61+-0.17\\
4 & STB & 2012-11-11 13:36 & 512.0 & 0.334 & 710 & 333+-22 & 1.33+-0.46 & 2012-11-12 21:27 & 436+-150 & 0.85+-0.29\\
5 & STB & 2012-11-19 09:50 & 505.0 & 0.326 & 643 & 344+-6 & 1.50+-0.54 & 2012-11-20 21:47 & 377+-137 & 0.75+-0.27\\
6 & STB & 2012-11-28 03:36 & 347.0 & 0.318 & 440 & 331+-7 & 1.46+-0.27 & 2012-11-29 14:32 & 378+-71 & 1.09+-0.20\\
7 & STA & 2013-05-29 12:20 & 480.0 & 0.516 & 879 & 398+-20 & 2.53+-0.37 & 2013-06-01 00:57 & 354+-51 & 0.74+-0.11\\
8 & STA & 2013-06-27 16:17 & 397.0 & 0.551 & 732 & 343+-20 & 2.53+-0.48 & 2013-06-30 04:54 & 377+-72 & 0.95+-0.18\\
9 & STA & 2013-07-25 06:12 & 545.0 & 0.584 & 1000 & 325+-23 & 2.57+-0.56 & 2013-07-27 19:50 & 393+-85 & 0.72+-0.16\\
10 & STA & 2013-08-10 15:00 & 453.0 & 0.603 & 375 & 367+-15 & 2.57+-0.24 & 2013-08-13 04:38 & 407+-38 & 0.90+-0.08\\
11 & EARTH & 2014-02-15 13:45 & 450 & 0.669 & 620 & 342+-10 & 2.14+-0.37 & 2014-02-17 17:07 & 541+-93 & 1.20+-0.21\\
12 & EARTH & 2014-04-05 19:00 & 500 & 0.624 & 450 & 419+-22 & 1.84+-0.72 & 2014-04-07 15:10 & 587+-230 & 1.17+-0.46\\
13 & EARTH & 2014-04-18 19:00 & 500 & 0.609 & 396 & 366+-30 & 2.40+-0.71 & 2014-04-21 04:32 & 440+-131 & 0.88+-0.26\\
14 & EARTH & 2016-03-20 07:00 & 430 & 0.602 &  & 432+-19 & 3.00+-0.61 & 2016-03-23 06:55 & 348+-71 & 0.81+-0.16\\
15 & EARTH & 2016-08-02 14:00 & 460 & 0.418 & 350 & 350+-18 & 2.40+-0.53 & 2016-08-04 23:32 & 302+-66 & 0.66+-0.14\\
\hline
Average & & & 474.6 & 0.4857656719584 & 642 & 355+-4 & 2.12+-0.13 & & 406+-27 & 0.86+-0.06\\
\hline
\end{tabular}

\end{sidewaystable*}

\begin{sidewaystable*}
    \centering
    \caption{Table of all the ICMEs examined in this study. This table supplements the data from Table \ref{tab:icmes1} 
    with the average acceleration values calculated using equation \ref{eq:acceleration}, the time 
    $t_{\SI{21.5}{\solarradii}}$ used for the ENLIL simulations, the travel time $T_\text{ENLIL}$ 
    calculated from the ENLIL model results and the travel time $T_\text{DBM}$ calculated using DBM by propagating the 
    ICME from \SI{1}{\AU} to Mars (section \ref{subsec:models}) using a drag parameter of $\Gamma = 
    \SI{0.1e-7}{\per\kilo\meter}$. The last column shows an estimation of the actual drag parameter for this event 
    between the Sun and \SI{1}{\AU} calculated using the observation values as described at the end of section 
    \ref{subsec:dbm}, where the average displayed in the bottom row is weighted using the inverse errors. The 
    acceleration values of event 10 and 15 are very small with absolute errors similar to the others, which makes the 
    errors of $\Gamma$ large. However, due to the weighted mean calculation, these two values only have a very small 
    influence on the mean $\Gamma$ value for all events shown in the last row. $\Gamma$ values for the 
    propagation between \SI{1}{\AU} and Mars are not shown, their uncertainties are so large that the values are not 
    meaningful. For Event 15, there is a negative $\Gamma$ value because the speed at \SI{1}{\AU} is larger than the 
    launch speed, which is very low both in the DONKI and CACTUS ICME catalogs.}
    \label{tab:icmes2}
    
    \begin{tabular}{@{\extracolsep{4pt}}cccS[table-format=1.2(2)]S[table-format=1.2(2)]S[table-format=1.2]cS[table-format=1.2(2)]S[table-format=1.1,round-mode=figures,round-precision=2]S[table-format=3.3(4)]@{}}
\hline
& \multicolumn{3}{c}{Repetition from Table \ref{tab:icmes1}} & \multicolumn{2}{c}{Acceleration} & \multicolumn{3}{c}{Model input and results} &\\\cline{2-4}\cline{5-6}\cline{7-9}\cline{10-10}
{ICME} & {Obs S/C} & {$t_{\SI{1}{AU}}$} & {$T_{\text{correl}}$} & {$a_{\SI{1}{\AU}, \text{Mars}}$} & {$a_{\text{Sun}, \SI{1}{\AU}}$} & {$t_{\SI{21.5}{\solarradii}}$} & {$T_\text{ENLIL}$} & {$T_{\text{DBM}}$} & {$\Gamma_{{\mbox{$\odot$}}-\SI{1}{\AU}}$}\\
{} & & {(UTC)} & {/ \si{\day}} & {/ \si{\meter\per\second\squared}} & {/ \si{\meter\per\second\squared}} & {(UTC)} & {/ \si{\day}} & {/ \si{\day}} & {/ \SI{e-7}{\per\kilo\meter}}\\
\hline
1 & STB & 2012-09-25 16:26 & 1.24+-0.23 & -3.2+-1.3 & -2.13 & 2012-09-23 18:58 & 1.03+-0.14 & 1.0 & 0.082+-0.004 \\
2 & STB & 2012-10-17 06:57 & 2.05+-0.33 & -0.60+-0.47 & -3.03 & 2012-10-14 08:45 & 1.54+-0.26 & 2.1 & 0.224+-0.006 \\
3 & STB & 2012-10-25 19:10 & 2.31+-0.66 & -1.68+-0.28 & -0.07 & 2012-10-21 03:59 & 1.28+-0.34 & 2.5 & 0.058+-0.006 \\
4 & STB & 2012-11-11 13:36 & 1.33+-0.46 & -1.3+-2.2 & -1.33 & 2012-11-08 07:21 & 1.33+-0.23 & 1.8 & 0.17+-0.03 \\
5 & STB & 2012-11-19 09:50 & 1.50+-0.54 & -2.0+-1.4 & -0.73 & 2012-11-16 06:32 & 0.94+-0.19 & 1.7 & 0.138+-0.007 \\
6 & STB & 2012-11-28 03:36 & 1.46+-0.27 & 0.5+-1.2 & -0.29 & 2012-11-23 17:17 & 1.46+-0.28 & 1.6 & 0.7+-0.2 \\
7 & STA & 2013-05-29 12:20 & 2.53+-0.37 & -1.16+-0.30 & -2.68 & 2013-05-26 22:58 & 1.37+-0.25 & 1.9 & 0.34+-0.05 \\
8 & STA & 2013-06-27 16:17 & 2.53+-0.48 & -0.18+-0.63 & -1.97 & 2013-06-24 08:08 & 1.97+-0.41 & 2.6 & 0.40+-0.07 \\
9 & STA & 2013-07-25 06:12 & 2.57+-0.56 & -1.37+-0.47 & -3.85 & 2013-07-22 09:55 & 1.97+-0.44 & 2.1 & 0.19+-0.02 \\
10 & STA & 2013-08-10 15:00 & 2.57+-0.24 & -0.42+-0.30 & 0.35 & 2013-08-07 02:53 & 2.57+-0.26 & 2.4 & -2+-1 \\
11 & EARTH & 2014-02-15 13:45 & 2.14+-0.37 & 1.0+-1.2 & -0.57 & 2014-02-12 18:51 & 2.01+-0.32 & 2.6 & 0.15+-0.02 \\
12 & EARTH & 2014-04-05 19:00 & 1.84+-0.72 & 1.1+-3.3 & -0.23 & 2014-04-02 00:19 & 2.53+-0.30 & 2.4 & 0.7+-0.6 \\
13 & EARTH & 2014-04-18 19:00 & 2.40+-0.71 & -0.6+-1.1 & -0.01 & 2014-04-14 19:44 & 2.35+-0.23 & 2.2 & 0.01+-0.01 \\
14 & EARTH & 2016-03-20 07:00 & 3.00+-0.61 & -0.63+-0.42 &  &  &  & 3.0 &  \\
15 & EARTH & 2016-08-02 14:00 & 2.40+-0.53 & -1.53+-0.30 & 0.14 & 2016-07-29 08:50 & 2.31+-0.24 & 3.1 & -0.5+-0.3 \\
\hline
Average & & & 2.12+-0.13 & -0.81+-0.33 & -1.17 & & 1.76+-0.08 & 2.2 & 0.13 \\
& & & & {$\sigma = \num{1.11}$} & {$\sigma = \num{1.29}$}\\
\hline
\end{tabular}

\end{sidewaystable*}

\subsection{Statistical analysis}

\begin{figure}[htb]
	\centering
	\includegraphics{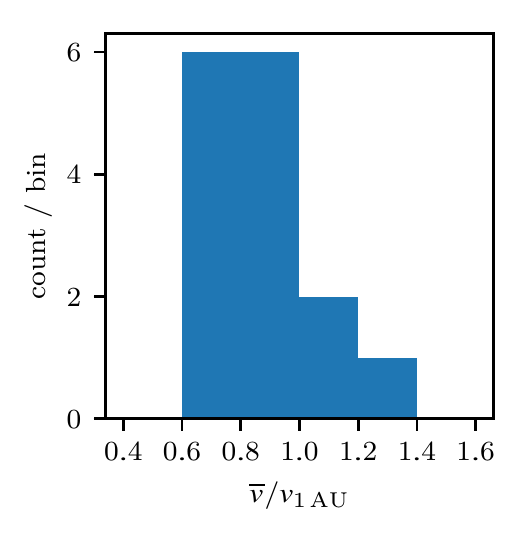}
	\caption{Histogram of ICME speed changes between \SI{1}{\AU} and Mars. Plotted is the ratio of the calculated mean 
	speed between \SI{1}{\AU} and Mars over the measured speed at \SI{1}{\AU}.}
	\label{fig:histogram}
\end{figure}

In Figure \ref{fig:histogram}, we show a histogram of the ratio $\overline{v}/v_{\SI{1}{\AU}}$ for the 15 ICMEs. On 
average, we get a value of
\[\left\langle\frac{\overline{v}}{v_{\SI{1}{\AU}}}\right\rangle = \num{0.86+-0.06},\]
which indicates that the average ICME in our sample decelerates slightly during its propagation between 
\SIrange[range-phrase={~and~}]{1}{1.5}{\AU}.

Considering the calculated standard deviations $\sigma$ of the $\overline{v}/v_{\SI{1}{\AU}}$ values (included in Table 
\ref{tab:icmes1}) and using a $1\sigma$ confidence interval, we can say that 8 ICMEs (\SI{53}{\percent} of our sample 
of 
ICMEs) decelerated ($\overline{v}/v_{\SI{1}{\AU}} + \sigma < 1$) and no ICME accelerated ($\overline{v}/v_{\SI{1}{\AU}} 
- \sigma > 1$) while the 7 remaining events showed neither a clear deceleration nor acceleration. We calculated the 
mean and standard deviation of $v_{\SI{1}{\AU}}$ of our 15 events to be \SI{466.9}{\kilo\meter\per\second} and 
\SI{84.5}{\kilo\meter\per\second} respectively, while the mean and standard deviation of $v_{\SI{1}{\AU}}$ of all ICMEs 
in the Richardson and Cane list from 2012 until 2016 (123 events) are 
\SI{489.2}{\kilo\meter\per\second} and \SI{114.2}{\kilo\meter\per\second}. Despite of the small sample of our events, 
the $v_{\SI{1}{\AU}}$ measurements seem 
to suggest that they are good in representing the average ICME speeds at \SI{1}{\AU}. However we still note that our 
derived probabilities of the changing of ICME speeds should be applied with caution because a) our accuracy is not high 
enough to find out the exact speed change of the remaining 7 events,  and b) the geometry of the ICME may affect our 
results, which will be discussed in more detail later (see also Figures \ref{fig:non_radial_propagation_cartoon} and 
\ref{fig:non_radial_propagation_enlil}).

\begin{figure}[htb]
    \centering
    \includegraphics{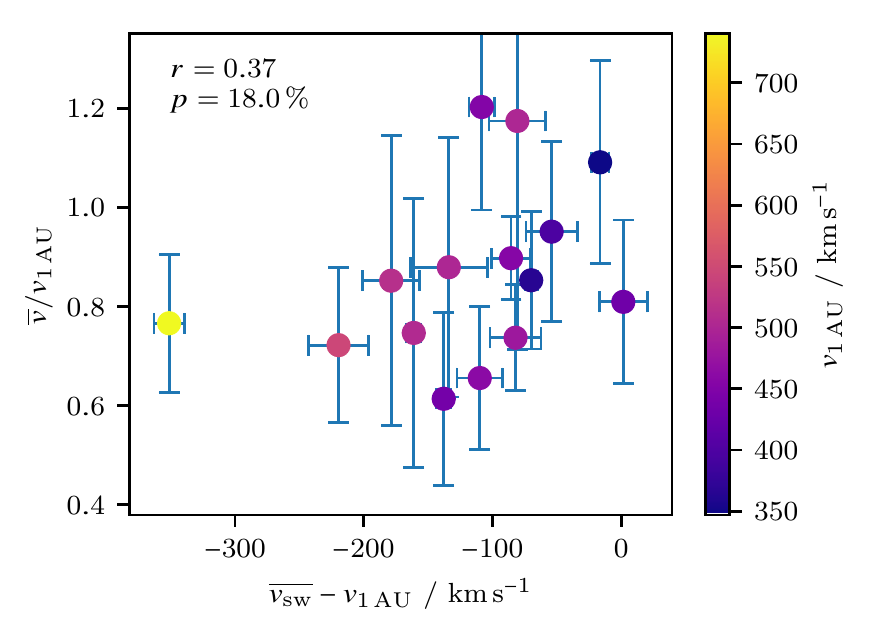}
    \caption{Comparison of the ratio $\overline{v}/v_{\SI{1}{\AU}}$ to $v_\text{sw} - v_{\SI{1}{\AU}}$, where 
    $v_\text{sw}$ is the ambient solar wind speed measured at ACE. The colors show the initial speed of the ICME at 
    $v_{\SI{1}{\AU}}$. The Pearson correlation coefficient $r$ and the probability $p$ that such a 
    dataset was produced by an uncorrelated system are displayed in the plot.}
    \label{fig:solarwind_scatter_plot}
\end{figure}

\begin{figure}[htb]
    \centering
    \includegraphics{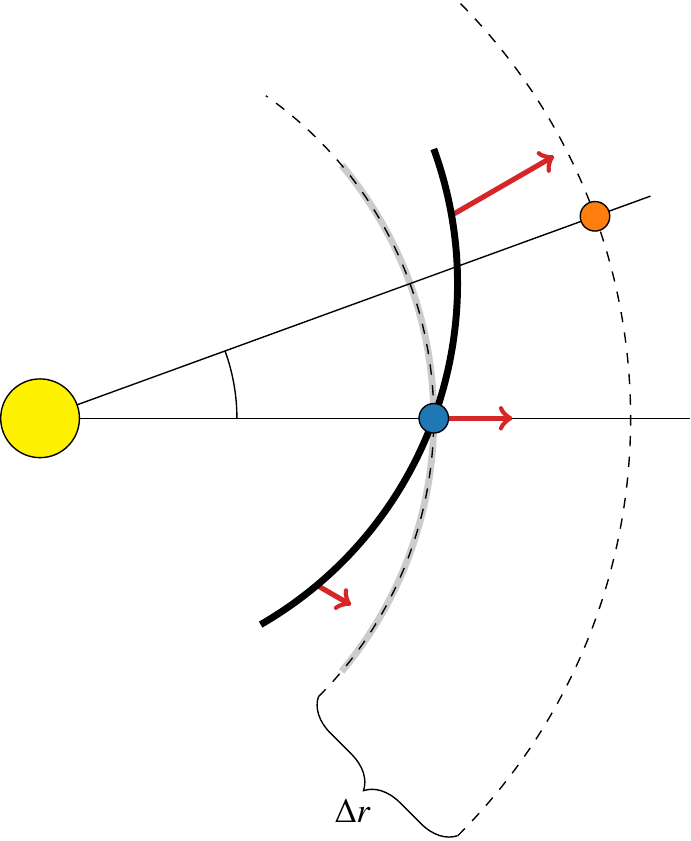}
    \caption{Cartoon illustration of the possible influence of the ICME shape on the measured speeds when the two 
    observation locations are not perfectly aligned in their heliospheric longitudes. In this case, the inclined shape 
    causes a percieved ``speedup'' of the ICME between Earth and Mars even if the actual speed of the ICME stays 
    constant.}
    \label{fig:non_radial_propagation_cartoon}
\end{figure}

\begin{figure}[htb]
    \centering
    \includegraphics[width=\linewidth]{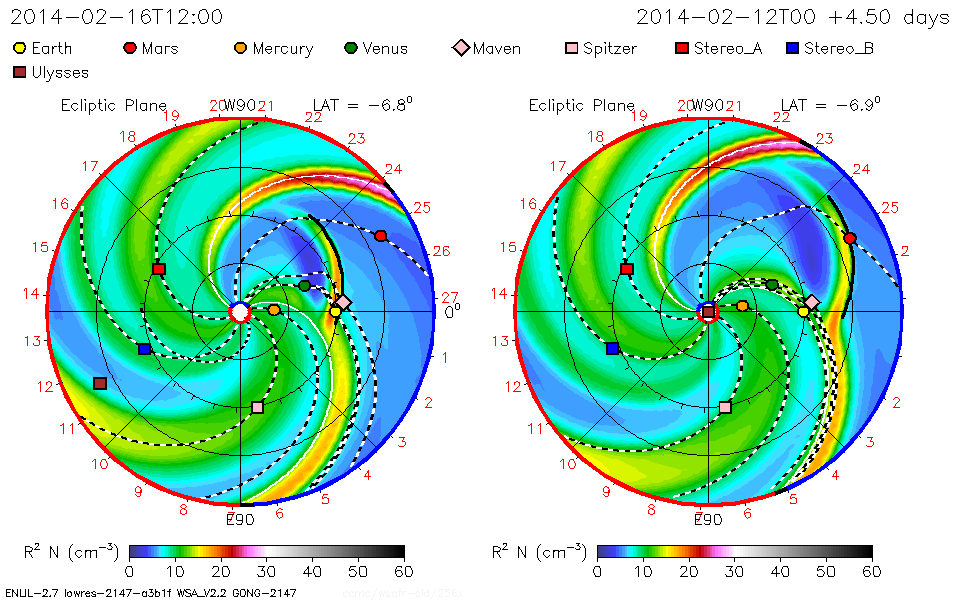}
    \caption{ENLIL simulation for the 2014-02-15 ICME, showing the same effect that was illustrated in Figure 
    \ref{fig:non_radial_propagation_cartoon} (left: arrival at Earth, right: arrival at 
        Mars). The CME front was emphasized manually using a black line.}
    \label{fig:non_radial_propagation_enlil}
\end{figure}

As the deceleration of ICMEs is believed to be related to the ambient solar wind speed, we also compared the 
$\overline{v}/v_{\SI{1}{\AU}}$ values to the solar wind speed $v_\text{SW}$ in Figure \ref{fig:solarwind_scatter_plot}, 
using data from the \textit{Solar Wind Electron, Proton, and Alpha Monitor} (SWEPAM) instrument \citep{McComas-1998} on 
the \textit{Advanced Composition Explorer} (ACE) spacecraft \citep{Stone-1998} located at the L1 
point near Earth and the \textit{Plasma and Suprathermal Ion Composition} (PLASTIC) \citep{Galvin-2008} instruments on 
the two STEREO spacecraft. The value we used for $v_\text{SW}$ 
is the average value of the solar wind speed measurements in a 1-day window before the 
ICME/disturbance arrival time at \SI{1}{\AU}, and its standard deviation was used for the error bars.

Most ICME speeds at \SI{1}{\AU} are larger than the ambient solar wind speed, which can be seen on the 
x axis in Figure \ref{fig:solarwind_scatter_plot}. Slightly different from previous findings, $\overline{v}$ (the 
average transit speed between \SI{1}{\AU} and Mars) is generally smaller 
than $v_{\SI{1}{\AU}}$ (which corresponds to a deceleration of the ICME), as visible on the y axis, apart from 3 cases 
where the error bars are also very large. Our results tend to show that lower ambient solar wind speeds compared to the 
ICME speed generally result in more deceleration even beyond \SI{1}{\AU}, as expected.

However, there is a considerable amount of variance in the data points, which is reflected by the Pearson correlation 
coefficient $r = \num{0.37}$ in Figure \ref{fig:solarwind_scatter_plot} not being very high.
This variance can possibly be due to the determined speed $\overline{v}$ being influenced by the 
geometry of the ICME: In general, the propagation of different parts of the ICME can be affected 
differently by ambient solar wind conditions and the interaction with other structures, such as stream 
interaction regions (SIRs)/corotating interaction regions (CIRs) and other ICMEs, potentially resulting 
in a variation of the ICMEs' geometric shape. This could lead to a radial asymmetry of the ICME 
resulting in larger uncertainties in our analysis especially when the two observers have a bigger longitudinal 
separation. A demonstration of this influence is also shown as a cartoon in Figure 
\ref{fig:non_radial_propagation_cartoon}, 
together with an example of the ENLIL model result in Figure \ref{fig:non_radial_propagation_enlil} (explained later in 
section \ref{subsec:models}) for ICME 11 where 
we suspect that this effect led to the ratio $\overline{v}/v_{\SI{1}{\AU}}$ being \num{1.20+-0.21}. Another example 
is visible in Figure \ref{fig:enlil-example}, where ICME 12 is merging with an SIR structure, possibly leading to a 
slight ``acceleration'', $\overline{v}/v_{\SI{1}{\AU}}=\num{1.17+-0.46}$. Similar effects have been observed 
previously by \citet{Prise-2015} and \citet{Winslow-2016}.

\begin{figure}[htb]
	\centering
	\includegraphics{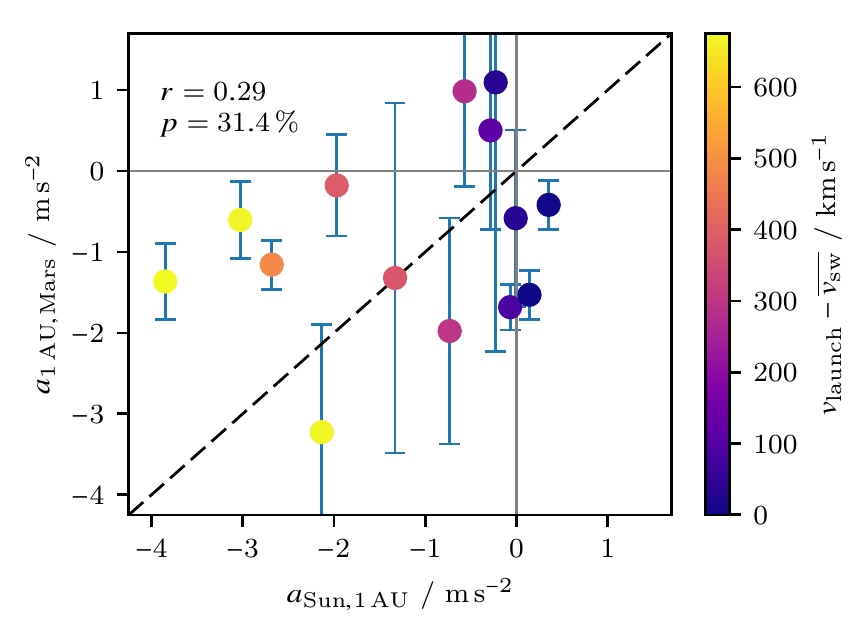}
	\caption{Comparison of the acceleration $a_{\text{Sun}, \SI{1}{\AU}}$ between \SI{21.5}{\solarradii} and the 
		arrival at \SI{1}{\AU} and $a_{\SI{1}{\AU}, \text{Mars}}$ between the arrival at \SI{1}{\AU} and the arrival at 
		Mars. The Pearson correlation coefficient $r$ and the probability $p$ that such a dataset was produced by an 
		uncorrelated system are displayed in the plot. The diagonal line marks where the accelerations would be equal 
		and the gray lines divide the four quadrants of the plot.}
	\label{fig:acceleration_scatter_plot}
\end{figure}

Another comparison can be made to the mean acceleration values that we calculated for the travel between 
\SI{21.5}{\solarradii} and \SI{1}{\AU} ($a_{Sun, \SI{1}{\AU}}$)  and between \SI{1}{\AU} and Mars ($a_{\SI{1}{\AU}, \text{Mars}}$) as shown in Figure 
\ref{fig:acceleration_scatter_plot}. The acceleration was calculated using equation \ref{eq:acceleration}, which 
depends on $T_\text{correl}^2$, amplifying the error bars. The big variations of $a$ shown in the figure 
indicate that ICMEs are very dynamic and their propagation depends on various properties, such as the different ambient 
solar wind conditions at different parts of the ICME and the interaction with other heliospheric structures. Our 
results suggest that the dynamics of the propagation continue to evolve beyond \SI{1}{\AU} and that, although the 
acceleration values up to and after \SI{1}{\AU} tend to be related (supported by a Pearson correlation coefficient of 
$r = \num{0.29}$, the acceleration is hardly a constant value. 
This is because \textit{a)} the ambient environment that the ICME travels through fluctuates due to the time-varying 
structures of the heliosphere \citep[as shown e.g. by][]{Temmer-2011} and \textit{b)} the ambient solar wind 
conditions vary throughout the heliosphere, thus diversely affecting the same ICME at different locations.

We have also marked the four quadrants in the plot, showing which ICMEs kept decelerating before and after 
\SI{1}{\AU} (lower left quadrant) and which changed from acceleration to deceleration (lower right) or the other way 
round (upper left). There are no ICMEs that accelerated before and after \SI{1}{\AU} (upper right quadrant) and two 
cases that ``accelerated'' between \SI{1}{\AU} and Mars were addressed above (event 11 and 12). There are also two 
ICMEs that seem to have accelerated between the Sun and \SI{1}{\AU}, i.e., event numbers 10 and 15, which have very 
low launch speeds reported (below \SI{400}{\kilo\meter\per\second} in the DONKI list, as well as even lower values in 
the SOHO/LASCO and CACTUS databases). These could of course be physical, but might also be due to the projection effect 
used in the image-based remote sensing analysis used to derive the launch speed. As the current paper is not focusing 
on the launch properties of the CMEs we did not pursue this matter further.

\begin{figure}[htb]
	\centering
	\includegraphics{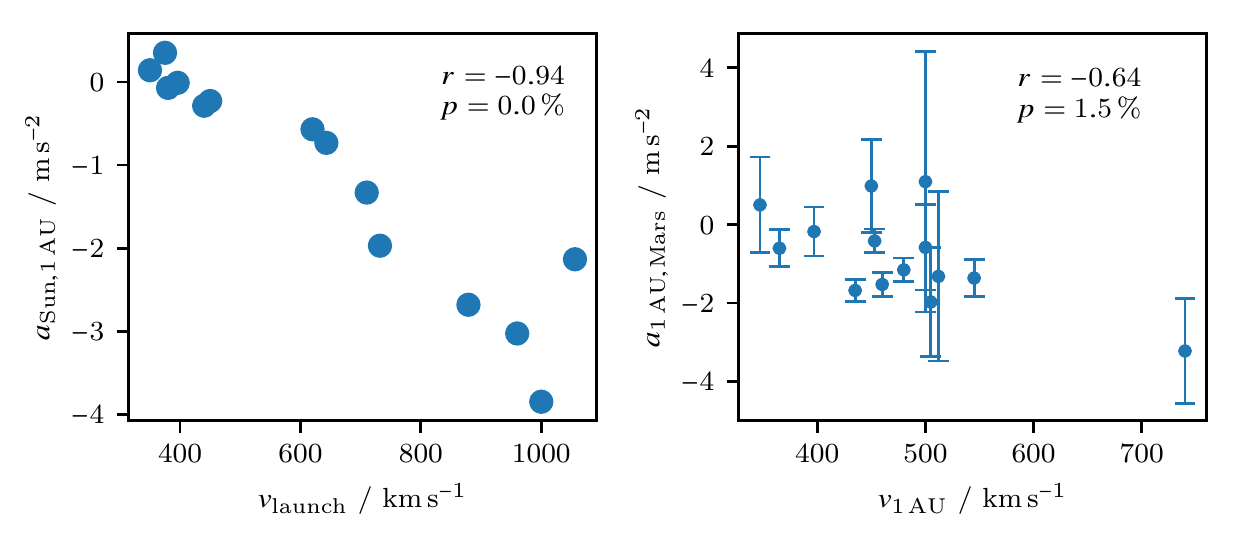}
	\caption{Comparison of the acceleration $a_{\SI{1}{\AU}, \text{Mars}}$ between the arrival at \SI{1}{\AU} and the 
	arrival 
	at Mars with the speed $v_{\SI{1}{\AU}}$ (right panel) and the acceleration $a_{\text{Sun}, \SI{1}{\AU}}$ between 
	\SI{21.5}{\solarradii} and the arrival at \SI{1}{\AU} with the speed $v_\text{launch}$ of the ICME at 
	\SI{21.5}{\solarradii} (left panel). The Pearson correlation coefficients $r$ and the probabilities $p$ that such 
	datasets were produced by an uncorrelated system are displayed in the plots.}
	\label{fig:acceleration_scatter_over_speed}
\end{figure}

In Figure \ref{fig:acceleration_scatter_over_speed}, we correlated the acceleration between the Sun and \SI{1}{\AU} 
with the launch speed (in the left panel) and the acceleration between \SI{1}{\AU} and Mars with the 
speed at \SI{1}{\AU} (in the right panel). Both plots show a stronger deceleration is correlated with 
higher ICME speeds, which 
is supported by high Pearson correlation coefficients of $r = \num{-0.94}$ and \num{-0.64} and low corresponding 
probabilities $p = \SI{0.0}{\percent}$ and \SI{1.5}{\percent} for uncorrelated data, respectively. Again, 
the error bars in the left panel are large due to the dependence of $a$ on $T_\text{correl}^2$.
Comparing our results for the acceleration with the values that \citet{Richardson-2014} obtained for ICMEs 
propagating from Earth to the Ulysses spacecraft (shown in their Figure 22 in a similar manner as our Figure 
\ref{fig:acceleration_scatter_over_speed}), which was at a distance of between 
\SIrange[range-phrase={~and~}]{3.74}{5.41}{\AU} from the Sun at that time, we find that our average deceleration value 
of \SI{0.81+-0.33}{\meter\per\second\squared} is much larger than their values of up to 
\SI{0.1}{\meter\per\second\squared}. This suggests that the deceleration becomes weaker at a larger radial distance 
beyond Mars, thus resulting in a lower average value between Earth and Ulysses.

\subsection{WSA-ENLIL+Cone model}
\label{subsec:models}

\begin{figure}[tbh]
    \centering
    \includegraphics[width=1\linewidth]{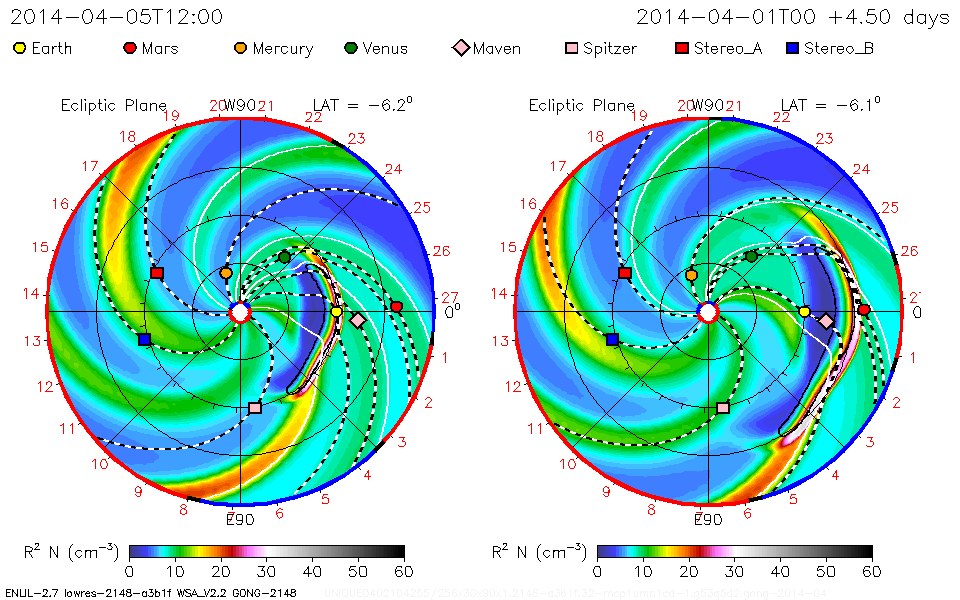}
    \caption{Example of an ENLIL simulation result for the 2014-04-05 ICME (left: arrival at Earth, right: arrival at 
        Mars).}
    \label{fig:enlil-example}
\end{figure}

The Wang-Sheeley-Arge (WSA) ENLIL model \citep{Odstrcil-2004} is a widely-used tool to predict solar wind 
propagation in the heliosphere. It is based on an MHD 
simulation and can be combined with a cone model to describe the propagation of ICMEs. Using the \textit{Space Weather 
Database Of Notifications, Knowledge, Information}\footnote{\texttt{https://kauai.ccmc.gsfc.nasa.gov/DONKI/}} (DONKI), which is 
based on coronagraph observations of CMEs close to 
their launch from the Sun, we matched most of the ICMEs in our study to WSA-ENLIL+Cone 
model simulation results provided by the CCMC \footnote{\texttt{https://ccmc.gsfc.nasa.gov/missionsupport/}}. In some cases, 
there 
are multiple ENLIL results for the same ICME (using slightly different input parameters) --- in that situation, we 
chose the one that gave the best \SI{1}{\AU} arrival time compared to the observations from the Richardson and Cane or 
Jian lists, respectively. In one case (ICME 14 in Table \ref{tab:icmes2}), we did not find any event output in the 
DONKI database that would possibly match the observed arrival time at \SI{1}{\AU}. Figure 
\ref{fig:enlil-example} shows a graphical representation of an ENLIL simulation result, specifically the ICME arriving 
at Earth on 2014-04-05 and at Mars on 2014-04-07, respectively (ICME 12 in Table \ref{tab:icmes2}).

To compare our measured ICME travel times to the ENLIL model results, we applied the cross-correlation method described 
in section \ref{subsec:cross-correlation} to the plasma number density $n$ at Earth (or STEREO) and Mars obtained 
from the model. This gives us another time lag value, which is considered to be the travel time that the ENLIL model 
predicts. In most cases, due to the smooth nature of the simulated data, the uncertainty of the travel time is smaller 
than for the one obtained from measured Forbush decreases.

\begin{figure}[htbp]
    \centering
    \includegraphics{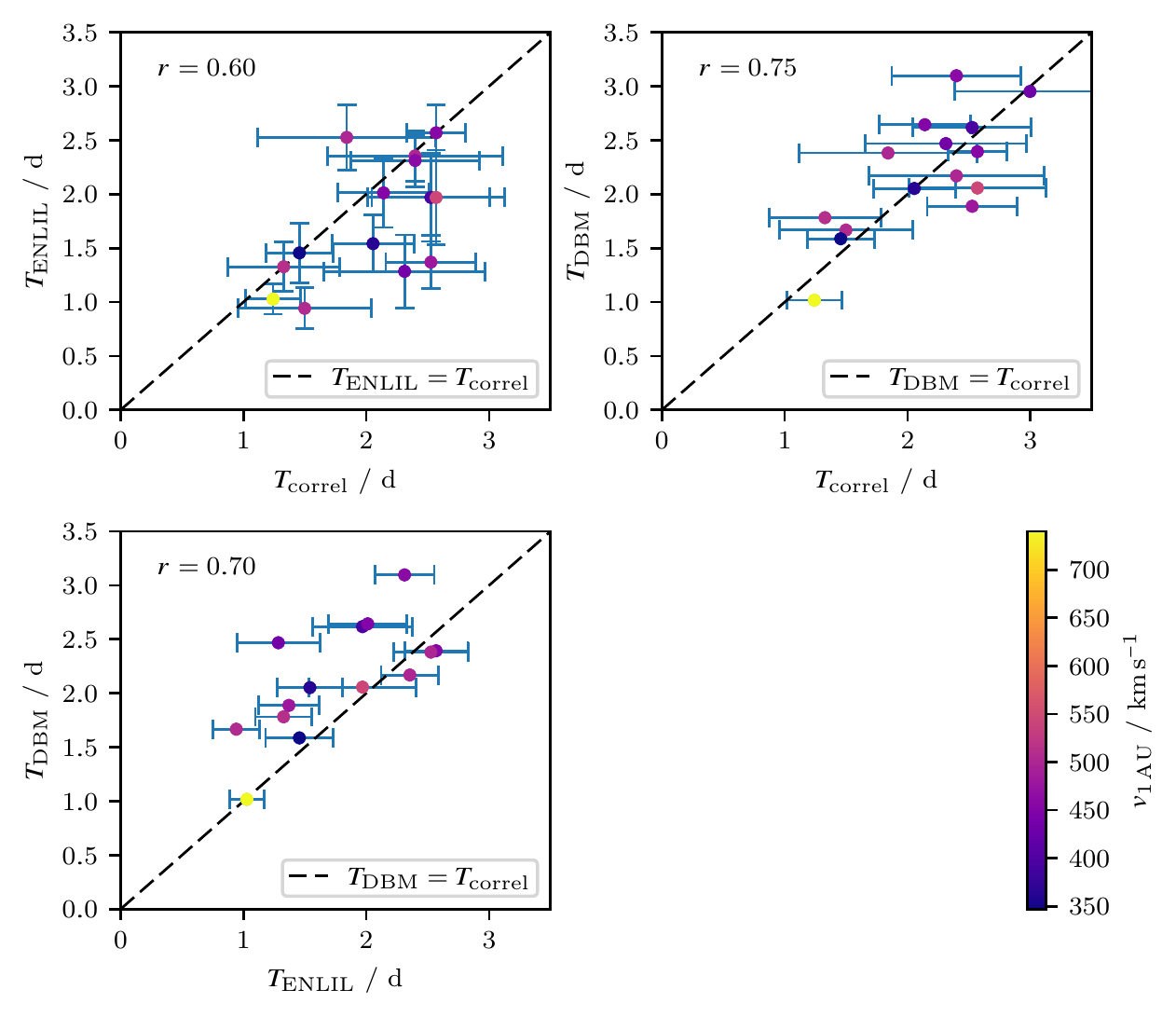}
    \caption{Plots comparing the ICME travel times between \SI{1}{\AU} and Mars determined using the cross-correlation 
    method and calculated by ENLIL or DBM. The diagonal line marks where the travel times would be equal. The Pearson 
    correlation coefficient $r$ is noted in the top left corner.}
    \label{fig:models_scatter_plot}
\end{figure}

In Figure \ref{fig:models_scatter_plot} (upper left panel), we compare the travel times calculated by the 
ENLIL model with the ones obtained 
from the \textit{in situ} data in this work. Both travel times are also listed in Tables \ref{tab:icmes1} and 
\ref{tab:icmes2}. For many 
events, ENLIL seems to predict a slightly faster propagation. 
Results for faster ICMEs (e.g. the 2012-11-28 ICME at STEREO B --- ICME 6 in Tables \ref{tab:icmes1} and 
\ref{tab:icmes2}) seem to agree 
quite well, while the slower events show larger differences. This might be the result of slower ICMEs being exposed to 
the disturbances in the interplanetary space for a longer time, thus accumulating a larger amount of possible 
uncertainties in the model. However, a more systematic statistical study based on more events should be carried out in 
the future to draw a solid statement on this matter.

We also calculated the mean difference between the results
\[\left\langle T_\text{ENLIL} - T_\text{correl} \right\rangle = \SI{-7+-11}{\hour}\]
and the average absolute difference
\[\left\langle \left|T_\text{ENLIL} - T_\text{correl}\right| \right\rangle \approx \SI{10}{\hour}.\]
The error given here is the standard error of the mean, not the standard deviation.

\subsection{Drag-Based Model}
\label{subsec:dbm}

A simpler model for the propagation of ICMEs is the Drag-Based Model (DBM), described in \citet{Vrsnak-2013} and 
\citet{Zic-2015}. The DBM is based on the assumption that beyond a distance of approximately $\SI{20}{\solarradii}$, 
the dominating influence on ICMEs is an ``aerodynamic'' drag force with an empirically determined drag parameter 
$\Gamma$. The main 
difference between DBM and ENLIL is that the former does not employ numerical MHD simulations --- the drag equations 
can be solved analytically. Therefore, the simulation is computationally inexpensive.

\citet{Vrsnak-2014} already compared results from DBM and ENLIL simulations and found that the ICME arrival times at 
Earth predicted by the two models generally agree quite well with average absolute-value difference of below 
\SI{8}{\hour}. For these results, drag parameter values between $\Gamma =$  
\SIrange[range-phrase={~and~}]{0.1e-7}{0.2e-7}{\per\kilo\meter} and solar wind speeds between $w =$ 
\SIrange[range-phrase={~and~}]{400}{500}{\kilo\meter\per\second} were used as an input for the DBM.

We apply the DBM model in such a way that the propagation of ICMEs is simulated starting from \SI{1}{\AU}, where the in 
situ measurement of the ICME is used as input, thus avoiding the uncertainty of the propagation from the Sun up to 
\SI{1}{\AU}. As the input for DBM, we used the local ICME speed ($v_{\SI{1}{\AU}}$) and the ambient solar wind speed 
for 
each event measured at ACE or STEREO as 
described in section \ref{sec:results}. The drag parameter $\Gamma$ was chosen to be \SI{0.1e-7}{\per\kilo\meter}, 
which is a low value that is commonly used for describing the propagation of the interplanetary shock 
associated with an 
ICME. We chose this value because the shock is related to the first step of the Forbush decrease 
\citep[e.g.][]{Cane-2000}. Assuming that the ICME propagates outward radially, the ICMEs' half-widths and the 
heliospheric longitudes of their propagation directions were taken from the DONKI database entries, as previously done 
for the ENLIL model (as such information is not available at \SI{1}{\AU}).

The arrival times at Mars predicted by DBM are marked in Figure \ref{fig:correlation_example} as well 
as Figures \ref{fig:icmes1} to \ref{fig:icmes5} in the appendix. Additionally, Figure 
\ref{fig:models_scatter_plot} (upper right panel) compares the travel times predicted by DBM to the results of the 
correlation method and the lower left panel compares the two models, ENLIL and DBM.

The mean difference and mean absolute difference between the results are:
\begin{align*}
\left\langle T_\text{DBM} - T_\text{correl} \right\rangle &= \SI{1+-9}{\hour}\\
\left\langle \left|T_\text{DBM} - T_\text{correl}\right| \right\rangle &\approx \SI{7}{\hour}\\
\left\langle T_\text{DBM} - T_\text{ENLIL} \right\rangle &= \SI{9+-10}{\hour}\\
\left\langle \left|T_\text{DBM} - T_\text{ENLIL}\right| \right\rangle &\approx \SI{11}{\hour}
\end{align*}
On average, DBM gives slightly better results than ENLIL for these events, even though the amount of variance is 
similar. Probably, this is primarily due to the fact that we could use DBM for propagation from \SI{1}{\AU} to Mars 
instead of from the Sun.

The agreement 
between the DBM and ENLIL models is similar to the one of ENLIL and the correlation method results with an average 
absolute-value difference slightly above the value of \SI{8}{\hour} determined by \citet{Vrsnak-2014} for the 
propagation up to \SI{1}{\AU}. This difference seems reasonable as the propagation from the Sun out to Mars ($\sim 
\SI{1.5}{\AU}$) takes a longer time and can therefore introduce a larger amount of error.

Under the assumption that the acceleration $a$ of an ICME stays constant between the Sun and \SI{1}{\AU} and between 
\SI{1}{\AU} and Mars and with a simplified, one-dimensional version of DBM (disregarding the influence of the geometric 
shape of the ICME), we also tried to derive the actual values of the drag parameter $\Gamma$ based on the 
observations and our calculated $a$ values by solving the following equation \citep[cf.][Equation 1]{Vrsnak-2013}
\begin{equation}
a = -\Gamma(v_\text{ICME}-v_\text{sw})|v_\text{ICME}-v_\text{sw}|
\end{equation}
for $\Gamma$.

Using the average speed $v_\text{ICME} = (v_\text{launch} + v_{\SI{1}{\AU}})/2$ and $a = a_{\text{Sun}, 
\SI{1}{\AU}}$ (from Table \ref{tab:icmes1} we get an estimation of $\Gamma$ between the Sun and \SI{1}{\AU}, which is 
given in Table \ref{tab:icmes2}. By calculating a weighted average using the inverse errors of these $\Gamma$ values, 
we obtain a result of \SI{0.09e-7}{\per\kilo\meter}, which shows that our assumption of $\Gamma = 
\SI{0.1e-7}{\per\kilo\meter}$ was reasonable. Nonetheless, we note that the variance of $\Gamma$ for different events 
is considerable which reflects the dynamic and variant nature of ICMEs and suggests that the approximated constant 
value of $\Gamma$ in DBM may result in uncertainties in the modelling procedure. The same  $\Gamma$ values could also 
be calculated between \SI{1}{\AU} and Mars 
using $v_\text{ICME} = \overline{v}$ and $a = a_{\SI{1}{\AU}, \text{Mars}}$, however, the results have propagated 
uncertainties that are too large to be meaningful.

\section{Conclusion}

We have described a method to determine the travel time of ICMEs between two heliospheric locations using the 
cross-correlation 
function of two \textit{in situ} data sets and applied it to 15 ICMEs and their Forbush decreases observed at Earth or 
the STEREO spacecraft and Mars close to their oppositions between 2012 and 2016. The method gives meaningful results 
in most cases apart from periods when ICMEs interact with each other and/or with SIRs/CIRs.

The results were used as the basis for this first statistical study of ICME-caused FDs observed at both \SI{1}{\AU} and 
\SI{1.5}{\AU}. It was found that the average ICME in our sample slightly decelerated during 
its propagation between \SI{1}{\AU} and \SI{1.5}{\AU}. Additionally, the results support that slower ambient solar 
wind speeds in comparison to the maximum ICME speed lead to a larger amount of deceleration. More studies based on a 
higher number of events in the future would help to better quantify these results.

The travel times between \SI{1}{\AU} and Mars obtained for the 15 events were compared with results from the ENLIL and 
DBM models. To derive travel times from the interplanetary plasma number density data output by ENLIL for different 
locations, the 
same cross-correlation method was used. On average, ENLIL predicts a faster propagation from \SI{1}{\AU} to Mars, but 
the ENLIL results seem to be less accurate for slower ICMEs in the study, which might be an effect of accumulation of 
uncertainties during the longer travel time.

Additionally, the observations were compared to results from the Drag-Based Model. Unlike ENLIL, 
we could use the observations at \SI{1}{\AU} as the basis for DBM and simulate the propagation 
from \SI{1}{\AU} to Mars. Avoiding the uncertainties of the propagation close to the Sun, this 
led to a slightly better agreement with the observations at Mars.

This highlights the importance of space weather modeling taking into account not only 
information about the launch of CMEs at the Sun, but also the in situ measurements further away, 
e.g. at \SI{1}{\AU}, to improve forecasts for space weather hazards for robotic missions positioned beyond \SI{1}{AU}. 
With future missions, such as Solar Orbiter and the Parker Solar Probe, we will have more measurements available 
at solar distances of less than \SI{0.3}{\AU}, which should be exploited as an input for modeling of space 
weather scenarios.

\acknowledgments
RAD is supported by NASA (HEOMD) under JPL subcontract \#1273039 to Southwest Research Institute and in Germany by 
DLR and DLR's Space Administration grant numbers 50QM0501, 50QM1201, and 50QM1701 to the Christian Albrechts 
University, Kiel.

Jingnan Guo and Robert Wimmer-Schweingruber acknowledge stimulating discussions with the ISSI team
``Radiation Interactions at Planetary Bodies'' and thank ISSI for its hospitality.

The research leading to these results has received funding from the European Union's Horizon 2020 research and 
innovation programme under the Marie Skłodowska-Curie grant agreement No 745782.

Lan K. Jian is supported by NSF grants AGS 1259549 and 1321493.

Bojan Vr\v{s}nak, Ja\v{s}a \v{C}alogovi\'{c} and Mateja Dumbovi\'{c} acknowledge financial support by the Croatian 
Science Foundation under project 6212 ``Solar and Stellar Variability''.

We acknowledge the NMDB database (www.nmdb.eu), founded under the European Union's FP7 programme (contract no. 
213007) for providing data. The data from South Pole neutron monitor is provided by the University of Delaware with 
support from the U.S. National Science Foundation under grant ANT-0838839.

Simulation results have been provided by the Community Coordinated Modeling Center at Goddard Space Flight Center 
through their archive of real-time simulations (\texttt{http://ccmc.gsfc.nasa.gov/missionsupport}). The CCMC is a 
multi-agency partnership between NASA, AFMC, AFOSR, AFRL, AFWA, NOAA, NSF and ONR. ENLIL with Cone Model
was developed by D. Odstrcil at George Mason University.

We thank the ACE SWEPAM instrument team and the ACE Science Center for providing the ACE data.

We acknowledge the STEREO IMPACT and PLASTIC teams (NASA Contracts NAS5-00132 and NAS5-00133) for the use of the 
solar wind plasma and magnetic field data.

\appendix
\section{Cross-correlation analysis plots for each event}

\begin{sidewaysfigure*}
	\centering
	\includegraphics{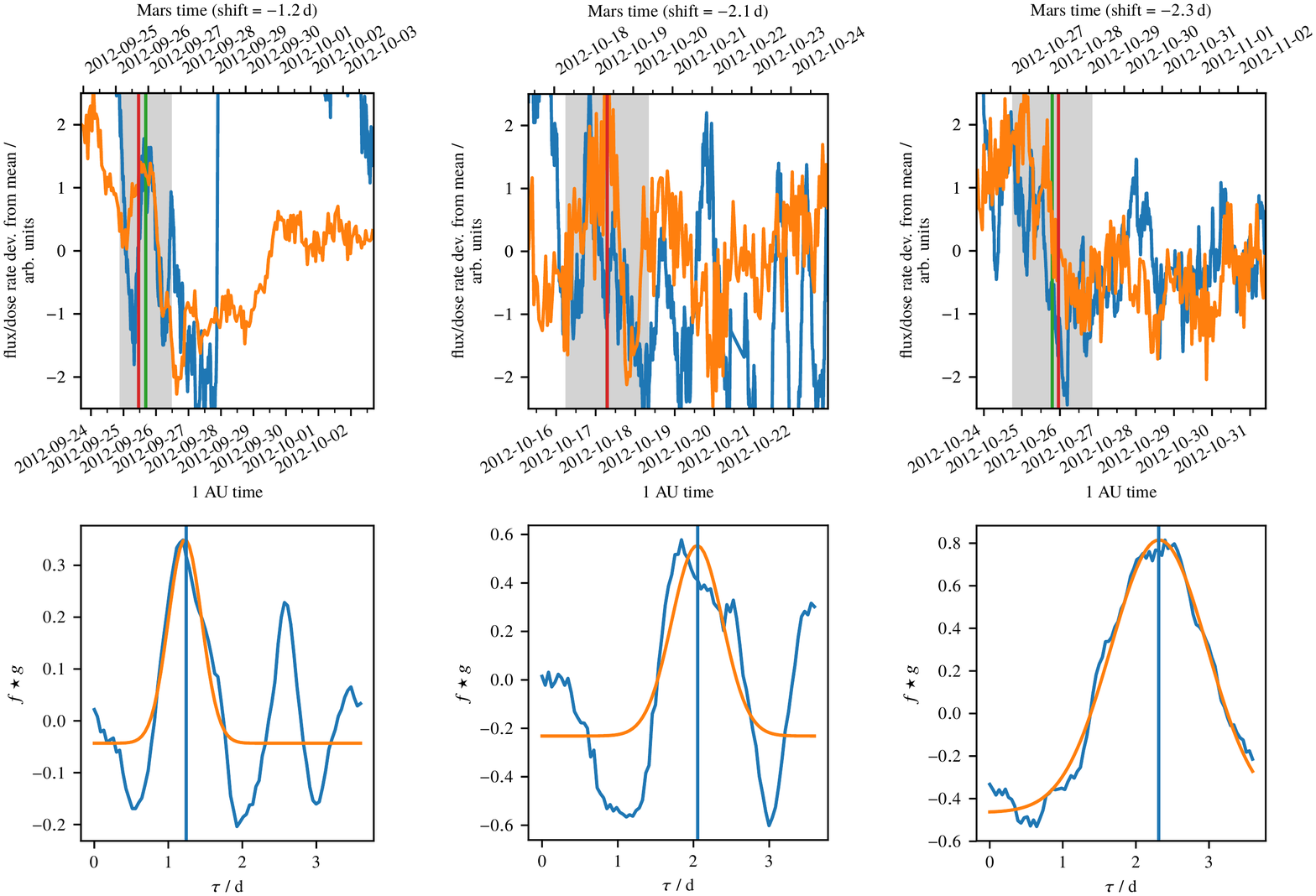}
	\caption{Plots showing the application of the cross-correlation method to every single ICME in the study. These are 
		Events 1 to 3 (all observed at STEREO B and Mars). The legend for the plots is in Figure 
		\ref{fig:correlation_example}.}
	\label{fig:icmes1}
\end{sidewaysfigure*}

\begin{sidewaysfigure*}
	\centering
	\includegraphics{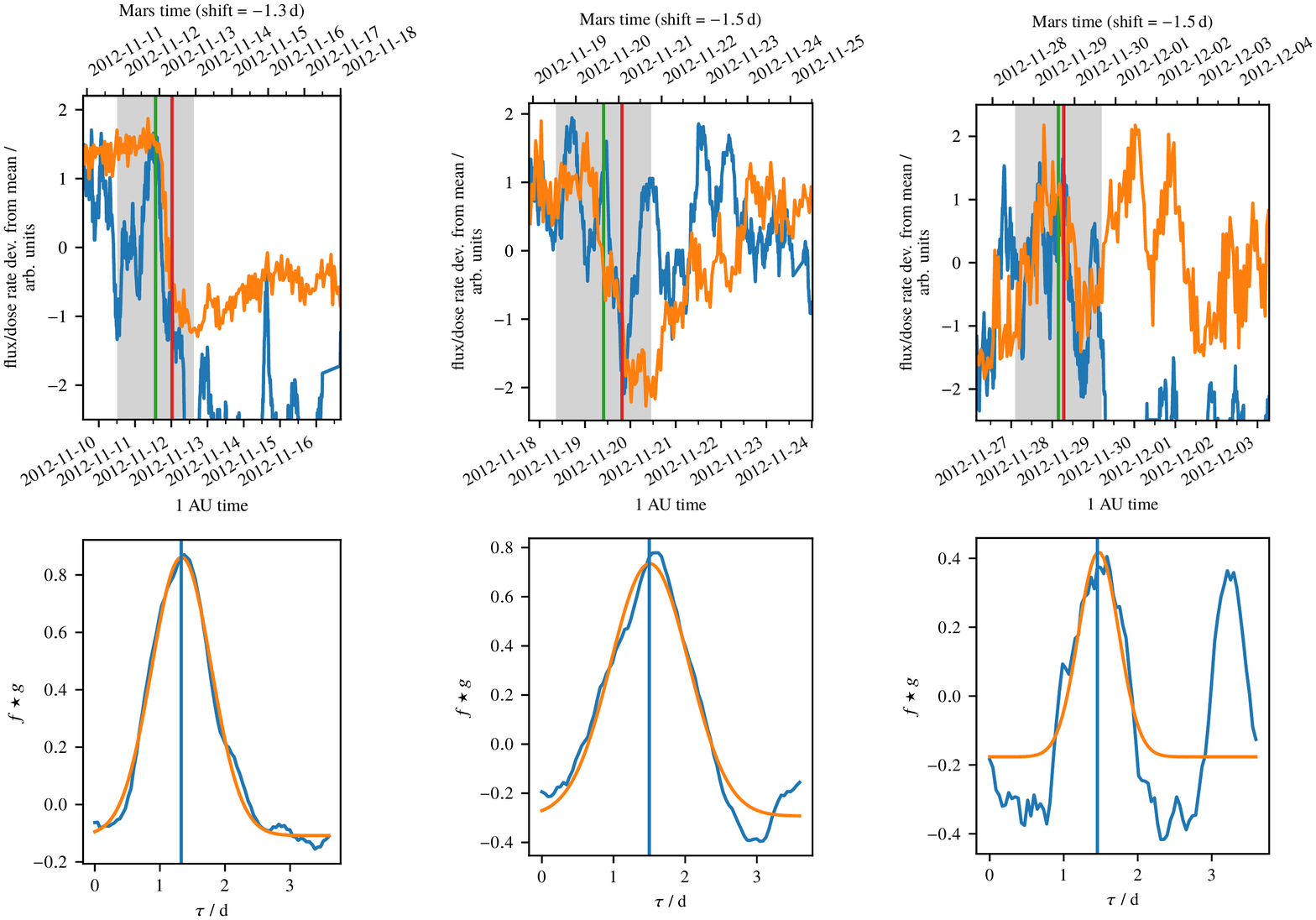}
	\caption{Plots showing the application of the cross-correlation method to every single ICME in the study. These are 
		Events 4 to 6 (all observed at STEREO B and Mars). The legend for the plots is in Figure 
		\ref{fig:correlation_example}.}
	\label{fig:icmes2}
\end{sidewaysfigure*}

\begin{sidewaysfigure*}
	\centering
	\includegraphics{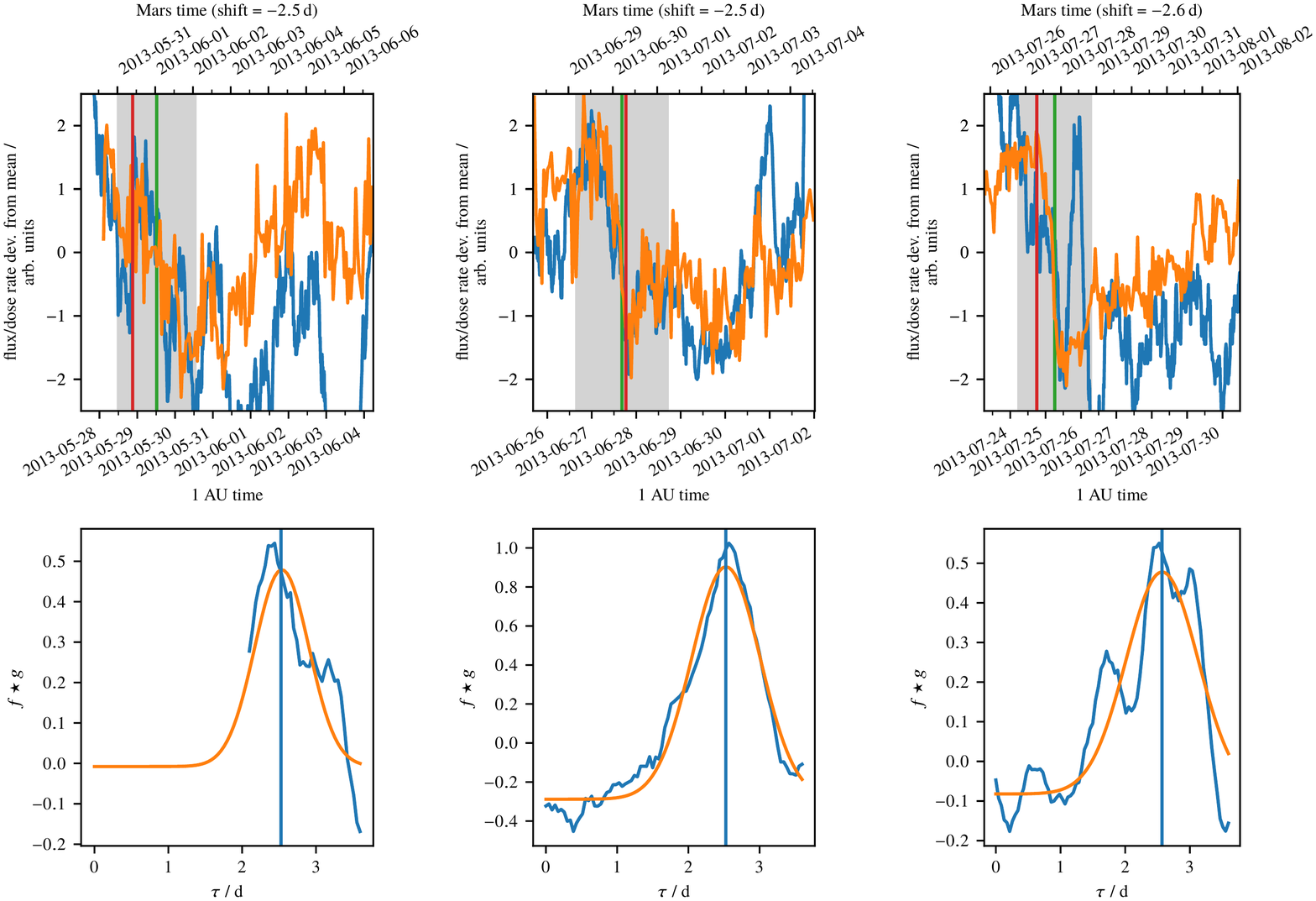}
	\caption{Plots showing the application of the cross-correlation method to every single ICME in the study. These are 
		Events 7 to 9 (all observed at STEREO A and Mars). The legend for the plots is in Figure 
		\ref{fig:correlation_example}.}
	\label{fig:icmes3}
\end{sidewaysfigure*}

\begin{sidewaysfigure*}
	\centering
	\includegraphics{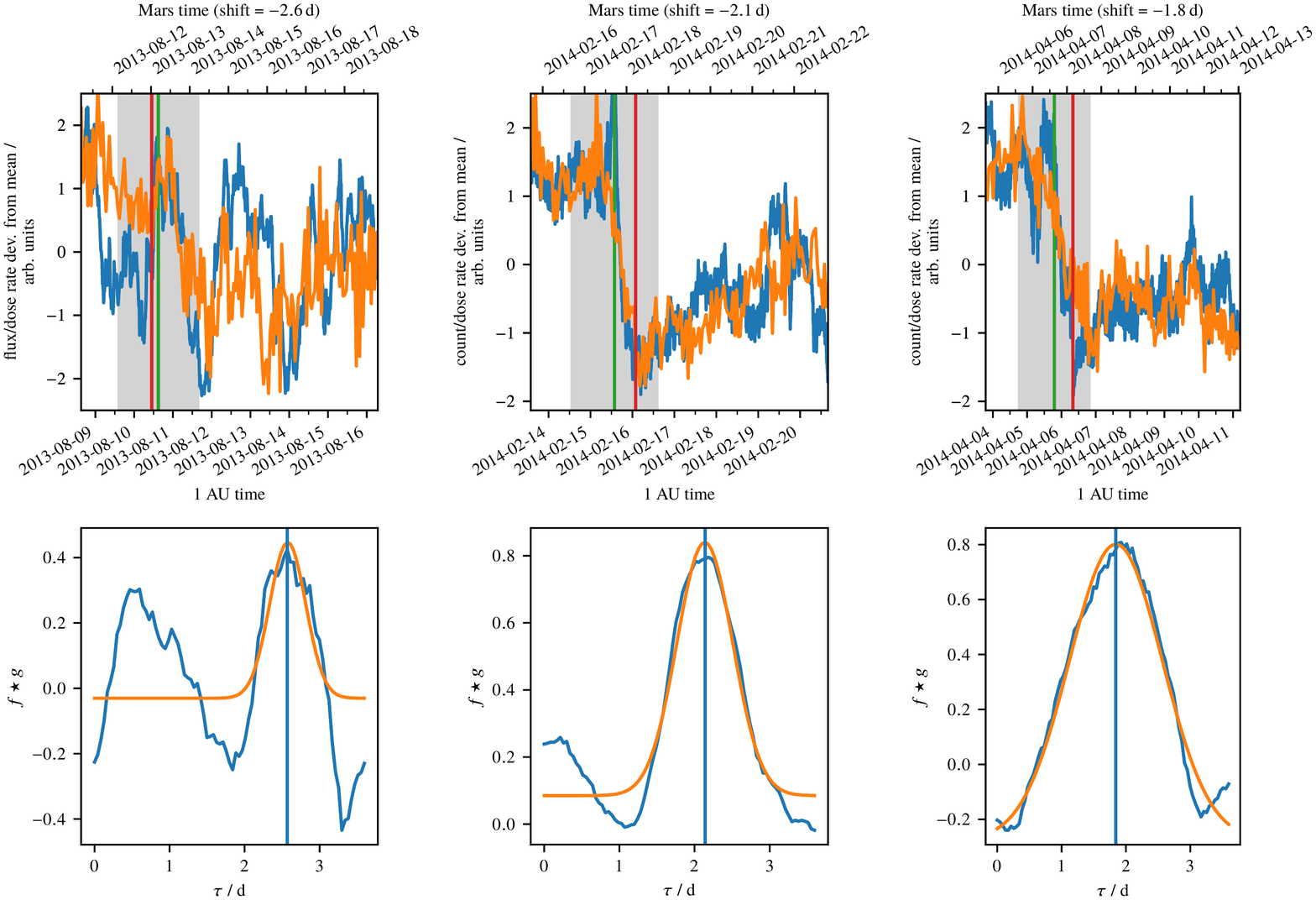}
	\caption{Plots showing the application of the cross-correlation method to every single ICME in the study. These are 
		Events 10 (observed at STEREO A and Mars) and 11 to 12 (observed at Earth and Mars).}
	\label{fig:icmes4}
\end{sidewaysfigure*}

\begin{sidewaysfigure*}
	\centering
	\includegraphics{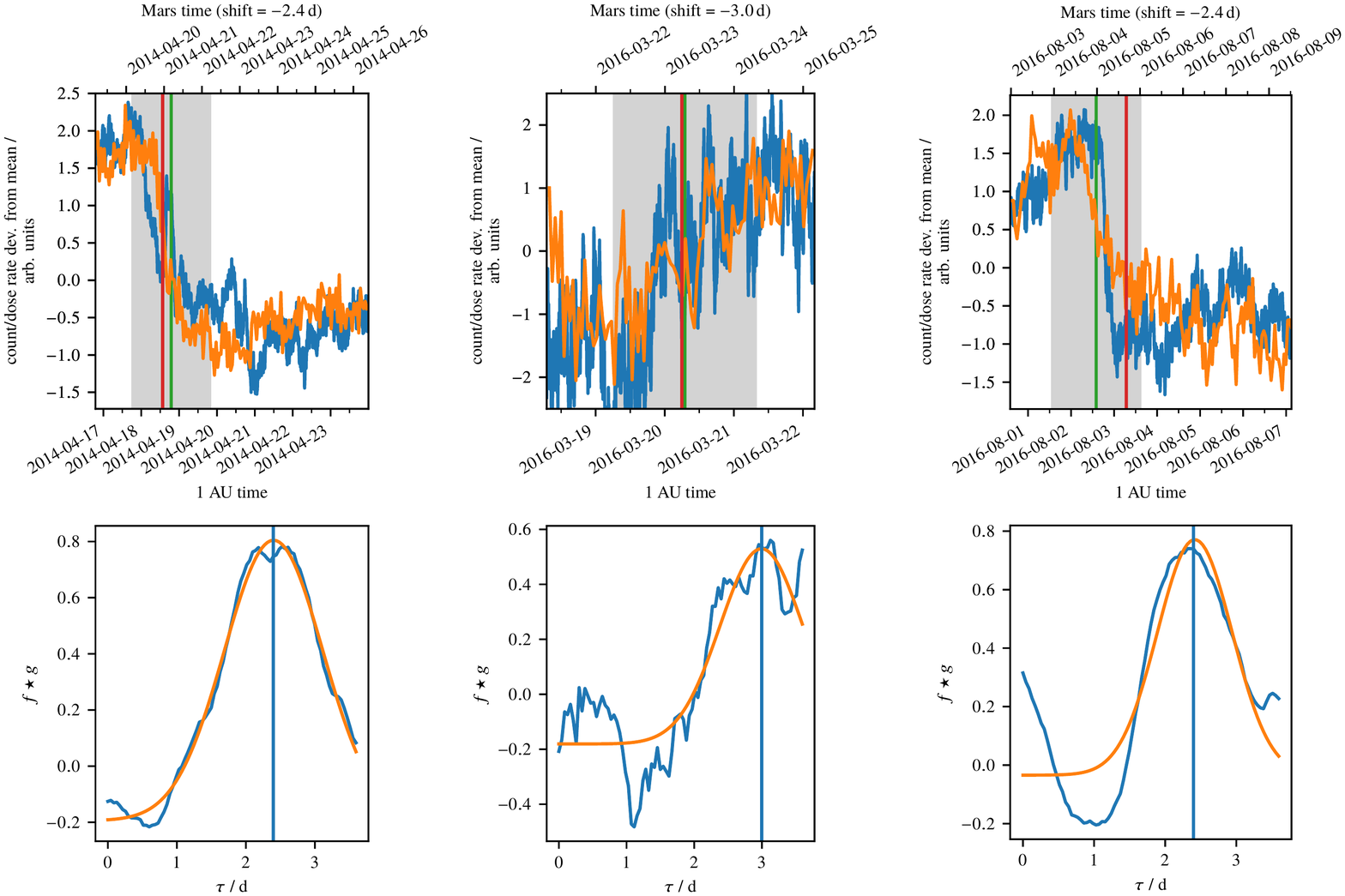}
	\caption{Plots showing the application of the cross-correlation method to every single ICME in the study. These are 
		Events 13 to 15 (all observed at Earth and Mars). The legend for the plots is in Figure 
		\ref{fig:correlation_example}.}
	\label{fig:icmes5}
\end{sidewaysfigure*}

\end{document}